\documentclass[12pt]{article}

\input{epsf}
\usepackage[nosort]{cite}
\usepackage{cite}
\usepackage{amsmath,pifont,bbding}
\usepackage{epsfig}
\usepackage{amsfonts}
\usepackage{amssymb}
\usepackage{multirow}
\usepackage{enumerate}
\usepackage{subfigure}
\usepackage{slashed}

\usepackage{color}

\usepackage{subfigure}

\usepackage{caption}

\newcommand{\be}{\begin{equation}}
\newcommand{\ee}{\end{equation}}
\newcommand{\bee}{\begin{equation*}}
\newcommand{\eee}{\end{equation*}}
\newcommand{\bea}{\begin{eqnarray}}
\newcommand{\eea}{\end{eqnarray}}
\newcommand{\bean}{\begin{eqnarray*}}
\newcommand{\eean}{\end{eqnarray*}}

\newcommand{\nn}{\nonumber}

\begin{document}

\setcounter{page}{0}
\thispagestyle{empty}

\begin{flushright}
DESY 16-157\\
\end{flushright}

\vskip 8pt

\begin{center}
{\bf \LARGE {Flavor Cosmology: Dynamical \\
\vskip 10pt
Yukawas in the  Froggatt--Nielsen Mechanism}}
\end{center}

\vskip 12pt

\begin{center}
 {\bf  Iason Baldes$^a$, Thomas Konstandin$^a$ and
G\'eraldine  Servant$^{a,b}$ }
 \end{center}

\vskip 14pt

\begin{center}
\centerline{$^{a}${\it DESY, Notkestra{\ss}e 85, D-22607 Hamburg, Germany}}
\centerline{$^{b}${\it II. Institute of Theoretical Physics, University of Hamburg, D-22761 Hamburg, Germany}}

\vskip .3cm
\centerline{\tt iason.baldes@desy.de, thomas.konstandin@desy.de, geraldine.servant@desy.de}
\end{center}

\vskip 10pt

\begin{abstract}
\vskip 3pt
\noindent

Can the cosmological dynamics responsible for settling down the present values of the Cabibbo-Kobayashi-Maskawa matrix be related to electroweak symmetry breaking? If the Standard Model Yukawa couplings varied in the early universe and started with order one values before electroweak symmetry breaking, the CP violation associated with the CKM matrix could be the origin of the matter-antimatter asymmetry. The large effective Yukawa couplings which lead to the enhanced CP violation can also help in achieving a strong first-order electroweak phase transition. We study in detail the feasibility of this idea by implementing dynamical Yukawa couplings in the context of the Froggatt--Nielsen mechanism. We discuss two main realizations of such a mechanism,  related phenomenology, cosmological and collider bounds, and provide an estimate of the baryonic yield. A generic prediction is that this scenario always features a new scalar field below the electroweak scale. 
We point out ways to get around this conclusion.

\end{abstract}

\newpage

\tableofcontents

\vskip 13pt

\newpage

\section{Introduction}

The origin of the highly hierarchical structure of the Standard Model Yukawa couplings remains an open question. This is the so-called flavor problem.
While many models and mechanisms have been advocated to explain this, the cosmological dynamics behind it has so far not been addressed.
This has several interesting ramifications. One first interesting question in this respect is to investigate the possible interplay with electroweak symmetry breaking.
Can the physics responsible for settling down the present values of the Cabibbo-Kobayashi-Maskawa (CKM) matrix be related to the dynamics of electroweak symmetry breaking?

There are many different aspects of this question. 
One is to check whether a model in which the new physics responsible for flavor arises at the TeV scale can  still be compatible with all experimental constraints. This is a direction which has already been vastly explored. The next step we want to take is to investigate whether the dynamics in the early universe at electroweak scale temperatures  can naturally lead to Yukawa couplings of order one before the EW phase transition. This question is of high interest for two reasons.
First it can lead to a first-order electroweak phase transition \cite{Baldes:2016rqn}. Second, it results in sufficient CP violation from the CKM matrix for electroweak (EW) baryogenesis \cite{dynamicalCKMCP}. 
As a result, 
the baryon asymmetry of the universe (BAU) can be generated during the electroweak phase transition (EWPT), in a process known as electroweak baryogenesis (EWBG).
In the Standard Model, EWBG fails. On the one hand, CP violation is suppressed by the quark masses and is way too small~\cite{Huet:1994jb,Gavela:1994dt,Brauner:2012gu}. On the other hand,  the Higgs is too massive, which results in a crossover transition~\cite{Kajantie:1996mn} with no departure from thermal equilibrium. 

Our motivation in this series of papers is to show that the interplay between Higgs and flavor dynamics  can naturally provide all conditions for successful EWBG. 
We already showed in Ref.~\cite{Baldes:2016rqn} in a model independent formulation that the variation of Standard Model Yukawa couplings induced during the EW  phase transition leads to drastic modifications of the phase transition. Yukawa couplings which decrease from $\sim \mathcal{O}(1)$ to their present values during the EWPT can create a thermal barrier between the symmetric and broken phase minima and hence make the EWPT strongly first-order. The origin of the BAU could therefore be closely linked to the flavor puzzle and the mechanism explaining the large hierarchy in the SM fermion masses. 
In fact, we also show in  Ref.~\cite{dynamicalCKMCP}
how the variation of SM Yukawa couplings during the EWPT naturally gives the correct amount of CP violation for EW baryogenesis.  A very interesting aspect of this framework is that it can circumvent usual CP violation limits. One generic  constraint on EW baryogenesis  is that any new source of CP violation needed to explain the BAU will typically be constrained by Electric Dipole Moment (EDM) bounds or leads to observable EDMs in the near future. In our scenario of time-dependent CP-violating source, we circumvent this usual constraint.\footnote{In the same spirit, EWBG using the SM  strong CP violation from the dynamical $\Theta_{QCD}$ parameter was advocated in \cite{Servant:2014bla}.}

We are therefore interested in exploring the model building conditions which favour such a scenario of variable CKM matrix during the EW phase transition. This should be done in the various classes of models which address the flavor problem.
Two main classes of models of flavor hierarchies are  Froggatt--Nielsen models and  Randall-Sundrum (RS)  models. 
This paper is about the Froggatt--Nielsen case. The RS case is presented in \cite{RSdynamicalCKMCP}.

In the Froggatt--Nielsen mechanism, the effective Yukawa couplings of the SM fermions depend on the vacuum expectation value (VEV) of an exotic scalar field~\cite{Froggatt:1978nt}. 
It was pointed out by Berkooz, Nir and Volansky that the CP violation associated with the CKM matrix could be large enough to explain the BAU if the VEV of the exotic Higgs --- and hence the effective Yukawa couplings --- were large at the start of the EWPT~\cite{Berkooz:2004kx}. However this scenario was not explored further.

The aim of this paper is to analyse this scenario, incorporating the dynamics of the Froggatt--Nielsen scalar field(s). We include the physics which actually leads to the variation of the Yukawas, which we ignored in~\cite{Baldes:2016rqn} and go through all experimental and cosmological constraints.

The order of the paper is as follows. In section \ref{sec:FNexplained} we introduce the Froggatt--Nielsen mechanism. In section \ref{sec:dynamics} we present two classes of models that realize the dynamics of Yukawa couplings at the EW scale. The details of each class of models are presented in Section \ref{sec:modelsA} and \ref{sec:modelsB}. We discuss the CP violation and baryonic yield in Section  \ref{sec:CP} and conclude in Section 7.  Some derivations are moved to the Appendix. 
In particular, experimental constraints are presented in Appendices \ref{section:flavorconstraints}
 and \ref{sec:constraints2}. We also argue in Appendix \ref{sec:dimension6} 
that our qualitative conclusions are expected to hold in general and do not depend on the specific form of the scalar potential which we chose for illustration and simplicity. 

\section{Froggatt--Nielsen mechanism}
\label{sec:FNexplained}
The Froggatt--Nielsen (FN) mechanism was proposed as a possible explanation for the mysterious hierarchy in the observed fermion masses~\cite{Froggatt:1978nt}. While there is a plethora of  implementations of this mechanism in the literature, the question of the dynamics and cosmology of this mechanism has not been addressed. In this paper, we want to study this question, in particular whether this dynamics could viably happen at the electroweak scale (while it is typically assumed that this  mechanism occurs at much higher scales). 

The main idea of the FN mechanism is that SM fermions carry different charges  under a new flavor symmetry, which is spontaneously broken by the VEV of a new scalar field $\langle S \rangle$, the so-called flavon field.  
Together with the flavon, a relatively large number of vector-like quarks are introduced to communicate the breaking of the flavor symmetry. The flavon field as well as all new quarks carry a Froggatt-Nielsen charge. The Yukawa couplings of the SM quarks are then generated by chain-like tree-level diagrams in which the difference in FN charge of the SM quarks is saturated by a number of insertions of the flavon field. Once the flavon obtains a VEV, the Yukawa couplings of the SM quarks are generated and are of order $(Y \langle S \rangle /M)^n$ where $n$ is the difference in FN charge, $Y$ is an order one coupling between the new quarks and the flavon and $M$ is the typical mass of a vector-like quark.
We define the scale of new physics as $\Lambda_s = M/Y$. 

The hierarchy in the SM Yukawa couplings is then given by the different powers of the ratio $\langle S \rangle/\Lambda_s$. We implement the mechanism with an exotic global symmetry $U(1)_{\rm FN}$ and a FN scalar $S$ which we choose, without loss of generality, to have charge $Q_{\rm FN}(S)=-1$ under the FN symmetry. 
We make a typical choice of FN charges for the SM quarks
	\begin{align}
	\overline{Q}_3 \; (0), \qquad \overline{Q}_2 \; (+2), \qquad \overline{Q}_1 \;(+3), \nonumber \\
	U_3 \; (0), \qquad U_2 \; (+1), \qquad U_1 \; (+4), \\
	D_3 \; (+2), \qquad D_2 \; (+2), \qquad D_1 \; (+3), \nonumber
	\end{align}
where $Q_{i}$ is the SM quark doublet, $U_{i}$ ($D_{i}$) is the right handed up (down) type quark and the subscripts denote generation~\cite{Berkooz:2004kx}. This gives  interactions of the form:
	\begin{align}
	\label{eq:FNyuks1}
	\mathcal{L} & = \tilde{y}_{ij}\left(\frac{S }{ \Lambda_{s} }\right)^{\tilde{n}_{ij}} \overline{Q}_i \tilde{\Phi} U_{j} + y_{ij}\left(\frac{S }{ \Lambda_{s} }\right)^{n_{ij}} \overline{Q}_i \Phi D_{j} \, ,
	\end{align}  
where $\Phi$ is the SM Higgs boson, $\tilde{\Phi}=i\sigma_{2}\Phi^{\ast}$, 
$y_{ij}$ are dimensionless couplings, assumed to be $\mathcal{O}(1)$ and $n_{ij}$ and $\tilde{n}_{ij}$
are chosen in such a way as to form singlets under $U(1)_{\rm FN}$.
The scalar $S$ obtains a VEV, $\langle S \rangle \equiv v_{s}/\sqrt{2}$, breaking the symmetry and resulting in Yukawa interactions between the SM Higgs and fermions.
 Defining
	\begin{equation} 
	\label{eq:epsdefined}
	\epsilon_{s} \equiv \frac{ \langle S \rangle }{ \Lambda_{s} } = \frac{v_{s}}{\sqrt{2}\Lambda_{s}} \, ,
	\end{equation}
one obtains Yukawa interactions of the form
	\begin{align}
	\mathcal{L} & =  \tilde{y_{ij}}\epsilon_{s}^{\tilde{n}_{ij}} \overline{Q}_i \tilde{\Phi} U_{j} + y_{ij}\epsilon_{s}^{n_{ij}} \overline{Q}_i \Phi D_{j} +H.c.
	\label{eq:FNyuks2}
	\end{align}
Fermion masses are generated when the SM Higgs gains a VEV, $\langle \Phi \rangle = (0 \; v_{\phi}/\sqrt{2})^{T}$ with $v_{\phi}=246$ GeV. Rotating to the mass basis for the quarks (see Appendix~\ref{sec:flavoncouplings}), the masses are related to the effective Yukawa couplings by $m_{f} = y^{f}_{\rm{eff}}v_{\phi}/\sqrt{2}$. For the FN charges above, one finds the effective Yukawa couplings to be
	\begin{align}
	y^{t}_{\rm{eff}} \sim 1, \qquad y^{c}_{\rm{eff}} \sim \epsilon_{s}^{3}, \qquad y^{u}_{\rm{eff}} \sim \epsilon_{s}^{7},  \nonumber \\
	y^{b}_{\rm{eff}} \sim \epsilon_{s}^{2}, \qquad y^{s}_{\rm{eff}} \sim \epsilon_{s}^{4}, \qquad y^{d}_{\rm{eff}} \sim \epsilon_{s}^{6}.
	\end{align}
Similarly one finds the pattern of CKM matrix elements
	\begin{equation}
	|V_{us}| \sim |V_{cd}| \sim \epsilon_{s}, \qquad |V_{cb}| \sim |V_{ts}| \sim \epsilon_{s}^{2}, \qquad |V_{ub}| \sim |V_{td}| \sim \epsilon_{s}^{3}.
	\end{equation}
The observed quark masses and mixing can therefore be accommodated with 
\be
\epsilon_{s} \sim 0.2  \ \mbox { today}.
\ee 
The explanation of the fermion mass hierarchy in the FN mechanism does not depend on the value of $\Lambda_s$, only on the ratio $\epsilon_s=\langle S \rangle/\Lambda_s$. If $\epsilon_s$ was of order 1 rather than 0.2 we would not observe any hierarchical structure and all Yukawas would be ${\cal O}(1)$. We are interested in the possibility that
\be
\epsilon_s \sim {\cal O}(1) \  \mbox{ before the EW phase transition,}
\ee
 as motivated by EWBG \cite{Baldes:2016rqn,dynamicalCKMCP}. As will be soon clear, this requirement imposes $\Lambda_s$ not to be much higher than the TeV scale. In this case, 
there are obviously experimental constraints on $\Lambda_s$ that we discuss in Appendix \ref{section:flavorconstraints}.
For reasons of simplicity,  in this paper, we restrict ourselves to the FN mechanism applied to the quark sector.

We now write the scalar fields as
	\begin{equation}
	\label{eq:recomponents}
	\Phi = \frac{1}{\sqrt{2}}\begin{pmatrix} G_{1} + i G_{2}  \\ \langle \phi \rangle + \phi + i G_{3} \end{pmatrix},  
	\qquad S = \frac{\langle \sigma \rangle +\sigma+   i \rho}{\sqrt{2}} \, ,
	\end{equation}
where the $G_{i}$ are the would be Goldstone bosons eaten by the $W$ and $Z$, $\langle \phi \rangle$ is the real constant Higgs background expectation value and $h$ is the SM Higgs field. Similarly, $\langle \sigma \rangle$ is the real constant flavon background and $\sigma$ and $\rho$ 
are the real and imaginary components of the flavon field  $S$.\footnote{To avoid notational clutter, from now on will always use $\sigma$ and $\phi$ whether we refer to the background value or the field excitation, but it should be clear from the context. The minimum of the potential at $T=0$ will be denoted by $(\phi,\sigma)=(v_{\phi},v_{s})$.} We denote their masses $m_{\sigma}$ and $m_{\rho}$ respectively.
If $\sigma$ is the only source of global $U(1)_{\rm FN}$ symmetry breaking, the associated Goldstone boson $\rho$ will lead to long range interactions between the quarks via the Yukawa couplings of eq.~(\ref{eq:FNyuks2}). We therefore assume explicit breaking of the $U(1)_{\rm FN}$ 
symmetry\footnote{The last term in (\ref{eq:explicitbreaking}) could for instance be generated from the $U(1)_{\rm FN}$ invariant terms
 $\mu_{SY}(Y^{\dagger}SS+YS^{\dagger}S^{\dagger}) $
where $Y$ is another scalar carrying charge $-2$ under $U(1)_{\rm FN}$  acquiring a VEV.} 
	\begin{equation}
	\label{eq:explicitbreaking}
	V(S) \supset -\mu_{s}^{2}S^{\dagger}S+\lambda_{s}(S^{\dagger}S)^{2}-A^{2}(SS+S^{\dagger}S^{\dagger}) \, .
	\end{equation}
Minimization of the above potential gives the relations
	\begin{equation}
	\label{eq:massiverho}
v_s^2\equiv \sigma_{min}^{2}=(\mu^2_s+2A^2)/\lambda_s,\ \ \ 	m_{\sigma}^{2}=2\mu_{s}^{2}+4A^{2}=2\lambda_{s}v_{s}^{2}, \ \ \quad m_{\rho}^{2}=4A^{2}.
	\end{equation}
In the above discussion we have ignored the contribution of $\Phi$, 
but this can trivially be included.
 We assume the mass hierarchy
 \be
  m_{\rho} > m_{\sigma} \, ,
  \label{eq:hierarchyrhosigma}
  \ee 
 which is viable provided $\mu_{s}^{2} <0$ in eq.~(\ref{eq:massiverho}). 
 We derive in Section \ref{section:flavorconstraints} constraints from flavor physics.
  Assuming (\ref{eq:hierarchyrhosigma}), the contributions of $\rho$ to the Wilson coefficients can be smaller than those of $\sigma$ and we derive constraints on $m_{\sigma}$ only. This is a conservative choice.  Assuming instead $m_{\rho} < m_{\sigma}$ (see \cite{Bauer:2016rxs}) would lead  to the same constraints on $m_{\rho}$ and, by construction, stronger constraints on $m_{\sigma}$, modulo some cancellations.
 The short summary of Appendix \ref{section:flavorconstraints}, see eqs.~(\ref{eq:epskaon}-\ref{eq:rc2sd}), is the typical constraint
\be
m_{\sigma} \Lambda_s \gtrsim (\mbox{a few TeV})^2 \, .
\label{mainconstraint}
\ee
As we mentioned earlier, $\Lambda_s$ is related to the mass scale of the new FN quarks which are a crucial part of the FN mechanism. 
In this paper, we do not include these new fermions in our analysis as they are not expected to play any relevant role. In fact, these FN fermions, which carry the same quantum numbers as the SM quarks, are vector-like.
The flavon VEV will enter the off-diagonal entries of the vector-like quark mass matrix, but this will not  change the overall FN scale $\Lambda_{s}$ as we only consider VEVs up to $\Lambda_s$ (in this range their $T=0$ one-loop corrections are not expected to be significant).
Besides, we consider temperatures $ T_{EW} \ll \Lambda_s$, hence their thermal effects are also negligible.
They will therefore play no role in the dynamics of the EW phase transition and do not introduce any new source of CP violation relevant for EW baryogenesis. On the other hand, searches at ATLAS and CMS put a bound closely approaching the TeV scale, $\Lambda_s \gtrsim 1$ TeV~\cite{Khachatryan:2015oba,Khachatryan:2015gza,Khachatryan:2015axa,Aad:2015kqa,ATLAS-CONF-2016-013,ATLAS-CONF-2016-072}.

 \section{Realizing the dynamics of Yukawa couplings at the electroweak scale}
 \label{sec:dynamics}

The  challenge for implementing a model of varying SM Yukawa couplings at the EW scale that leads to successful EW baryogenesis comes from the tension between two requirements. On the one hand, we need the VEV of the flavon to vary from a value of order ${\cal O}(\Lambda_s)$ to a value of order ${\cal O}(\Lambda_s/5)$ during the EWPT, as to induce the Yukawa coupling variation from values of order one to their present values. 
As we will see, this implies, for a typical two-field scalar potential, a small flavon mass, below the EW scale.
On the other hand, in the simplest minimal flavon model,  flavor constraints  such as (\ref{mainconstraint}) impose the flavon to be  in the multi TeV range, as shown explicitly  in Appendix \ref{section:flavorconstraints}.
This tension can be evaded if the flavon is super light, below 100 eV, however this requires a severe tuning of the quartic coupling between the Higgs and the flavon.
We are therefore led to consider slightly less simplistic models and we will show that rather minimal models may display the desired properties.

\begin{itemize}
\item {\bf Models A.}
The first model (A-1) we consider for pedagogical reasons is similar to the one proposed in \cite{Berkooz:2004kx}. There is a single FN scalar $S$, with the mass of its real component, $\sigma$, below the Kaon mass, $m_{\sigma} < M_{K}$. The combination of flavor and cosmological constraints is satisfied for $\Lambda_{s} \gtrsim 10^{12}$ GeV. The VEV $\sigma$ is close to the mediator scale $\Lambda_{s}$ before the phase transition. The Yukawa couplings of order unity before the phase transition will give the required CP violation and lead to the strong first-order phase transition. The fields will change in the following way during the electroweak phase transition
\begin{equation}
\phi: 0 \to v_{\phi} \qquad \sigma: \Lambda_{s} \to \Lambda_{s}/5. 
\end{equation}
The steps of the phase transitions are illustrated in figure~\ref{fig:sketch}.
\begin{figure}[t]
\begin{center}
\includegraphics[width=200pt]{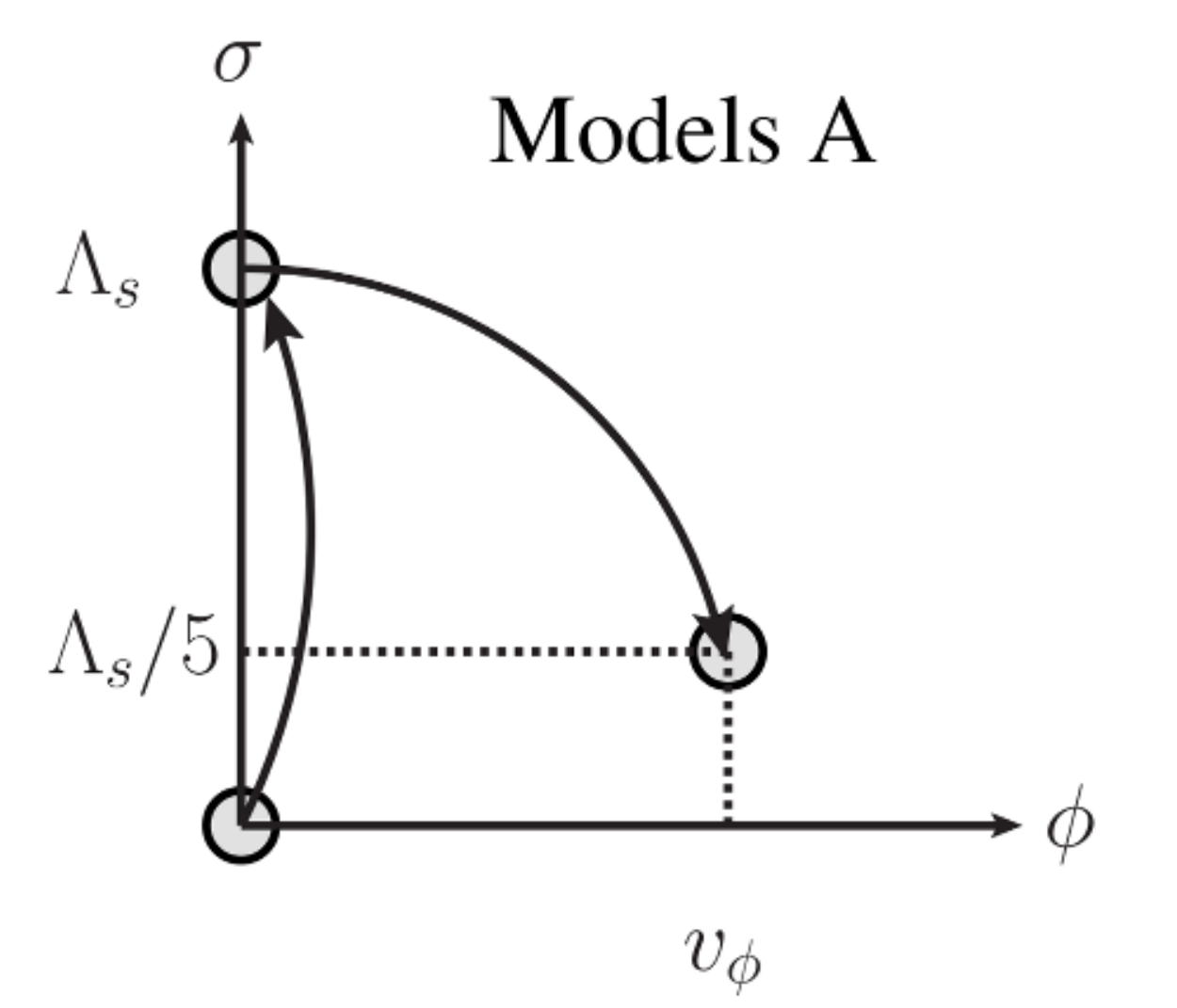}
\includegraphics[width=200pt]{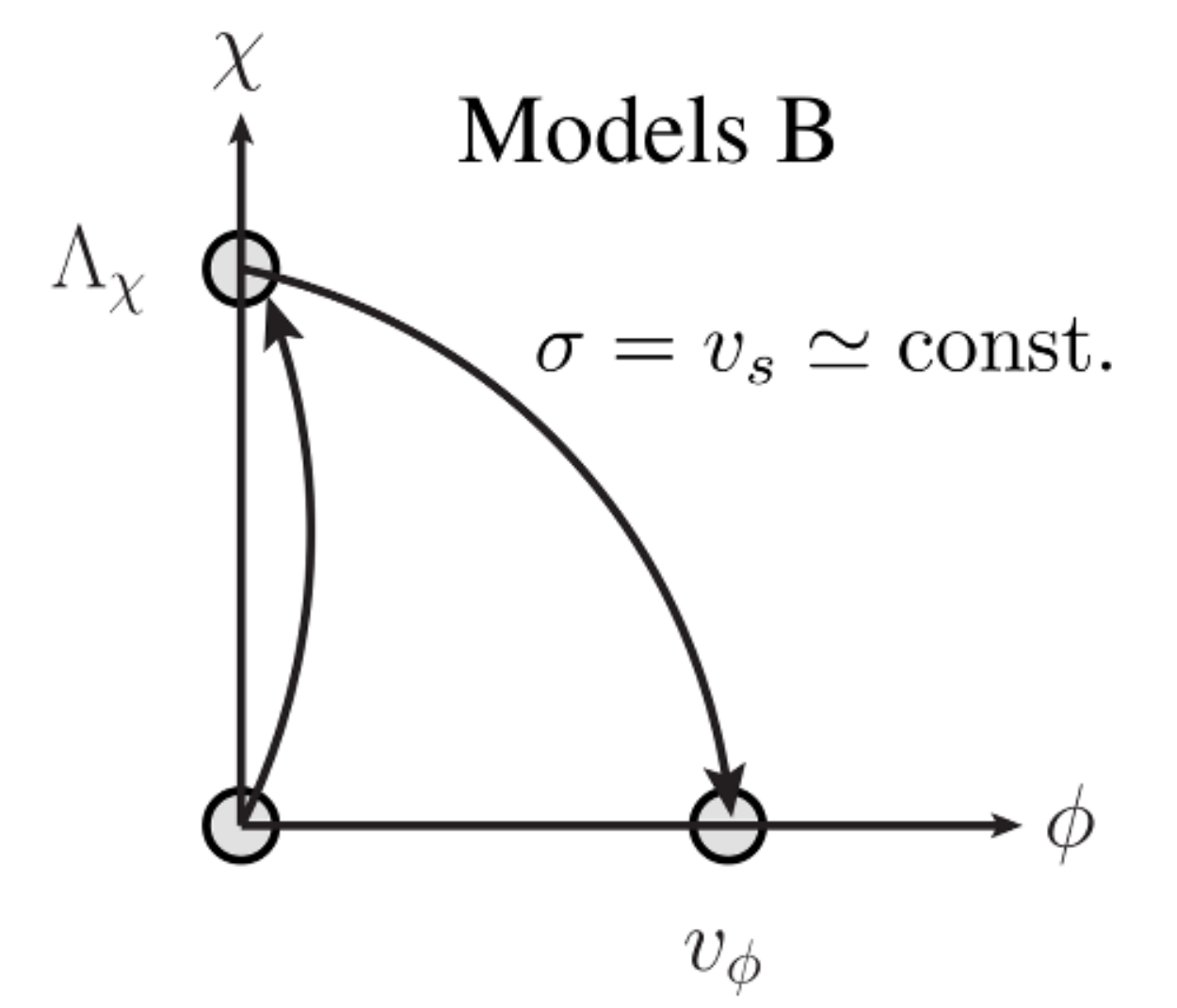}
\end{center}
\caption{\small  The behaviour of the fields for the two main realizations of the mechanism. The flavor symmetry first breaks to a minimum with large effective Yukawa couplings. The VEV of (one of) the flavons decreases during the subsequent EWPT, suppressing the Yukawa couplings.}
\label{fig:sketch}
\end{figure}
This minimal model leads to ridiculously small quartic couplings for the flavon, not stable under radiative corrections.
We therefore  consider some variation of this model, A-2, inspired  by the construction of Ref.~\cite{Knapen:2015hia} which enables to evade  flavor constraints and therefore to bring $\Lambda_s$ down to the TeV scale with unsuppressed quartic couplings. 

 \item {\bf Models B}

Since the difficulties in Models A come from the flavor constraints, we next consider a model with two FN scalars where only one gets a VEV today while the second one acquires a VEV at early times but has no VEV today. In this case, the first field $S$ (with real component  $\sigma$) is responsible for today's Yukawas while the second one $X$ (its real component  being denoted $\chi$) provides the CP violation needed for EW baryogenesis through its Yukawa couplings.
 The VEV $\chi$ is close to the mediator scale $\Lambda_{\chi}$ before the phase transition which leads to Yukawa couplings of order unity before the EWPT. 
 
 Generally, it is not easy in a one-field model to devise a potential that leads to symmetry  restoration at low temperatures. However, it is relatively easy in models with two scalar fields that display a two-stage phase transition. The simplest way to realize this in the current context is to couple the field $\chi$ to the Higgs while the VEV of the field $\sigma$ hardly changes. During the EW phase transition, the fields change in the approximate following way
\begin{equation}
\phi: 0 \to v_{\phi} \qquad \sigma: \Lambda_{s}/5 \to \Lambda_{s}/5 \qquad \chi: \Lambda_{\chi} \to 0. 
\end{equation}
We will consider the two cases corresponding to two choices of FN charges for $X$, $Q_{\rm FN}(X)=-1/2$ (model B-1) and  $Q_{\rm FN}(X)=-1$ (model B-2).

\end{itemize}

In summary, models B-1 and B-2 are the two viable examples we provide in the FN framework, realizing the possibility of varying quark Yukawa couplings during the electroweak phase transition, from values of order 1 in the EW symmetric phase to small values in the EW broken phase.
We give the details below on how this works as well as phenomenological bounds and signatures associated with these models.

\section{Models A: A single flavon}
\label{sec:modelsA}
\subsection{Tree level potential}
\label{sec:treelevelmodelA}
We begin with a renormalizable tree-level potential for $\Phi$ and $S$ given by
	\begin{align}
\label{eq:PotmodelA}
	V  = \mu_{\phi}^{2}\Phi^{\dagger}\Phi+\lambda_{\phi}(\Phi^{\dagger}\Phi)^{2} + \mu_{s}^{2}S^{\dagger}S+\lambda_{s}(S^{\dagger}S)^{2} + \lambda_{\phi s}(\Phi^{\dagger}\Phi)(S^{\dagger}S). 
	   	\end{align}
In terms of the dynamical field components the potential becomes
	\begin{equation}
	\label{eq:modelAtree}
	V=\frac{\mu_{\phi}^{2}}{2}\phi^{2}+\frac{\lambda_{\phi}}{4}\phi^{4}+\frac{\lambda_{\phi s}}{4}\phi^{2}\sigma^{2}+\frac{\mu_{s}^{2}}{2}\sigma^{2}+\frac{\lambda_{s}}{4}\sigma^{4}.
	\end{equation}
It is convenient to rewrite the above potential in the form
	\begin{equation}
	V=\frac{\mu_{\phi}^{2}}{2}\phi^{2}+\frac{\lambda_{\phi}}{4}\phi^{4}+\frac{\lambda_{s}}{4}\left(\sigma^{2}-\Lambda_{s}^{2}\left[1-C\frac{\phi^{2}}{v_{\phi}^{2}}\right]\right)^{2}-\frac{\lambda_{s}\Lambda_{s}^{4}}{4}\left[1-C\frac{\phi^{2}}{v_{\phi}^{2}}\right]^{2},
	\end{equation}
where $C$ is dimensionless. The parameters are related by
	\begin{align}
	\mu_{s}^{2}&=-\lambda_{s}\Lambda_{s}^{2}, \\
	\lambda_{\phi s}&=\frac{2C\Lambda_{s}^{2}}{v_{\phi}^{2}}\lambda_{s}. \label{eq:parrel}
	\end{align}
From the latter form of the potential one sees that for $\phi=0$ there is a minimum at $\sigma=\Lambda_{s}$, while for $\phi=v_{\phi}$ there is a minimum at $\sigma=\Lambda_{s}\sqrt{1-C}\equiv v_{s}$. In the baryogenesis mechanism discussed above, we require $\sigma$ to move from $\Lambda_{s}$ to approximately $\sqrt{2}\Lambda_{s}/5$ during the electroweak phase transition. To achieve this simply requires $C\approx 0.92$. 

The above Lagrangian contains five parameters. However, after setting the Higgs mass $m_{\phi}=125$ GeV, the electroweak VEV $v_{\phi}=246$ GeV and $C\approx0.92$ as discussed above, we are left with two free parameters which we take to be $\lambda_{s}$ and $\Lambda_s$. 

Stability of the tree level potential requires
	\begin{equation}
	\lambda_{\phi} > 0, \quad \lambda_{s} > 0, \quad \lambda_{\phi s} > -2\sqrt{\lambda_{s}\lambda_{\phi}} \, ,
	\end{equation}
while a minimum at $v_{\phi}\neq0$ ($v_{s}\neq0$) requires $\mu_{\phi}^{2} < 0$ ($\mu_{s}^{2} < 0$). The VEVs are related by
	\begin{align}
	\mu_{\phi}^{2}+\lambda_{\phi}v_{\phi}^{2}+\frac{\lambda_{\phi s}}{2}v_{s}^{2} = 0, \\
	\mu_{s}^{2}+\lambda_{s}v_{s}^{2}+\frac{\lambda_{\phi s}}{2}v_{\phi}^{2} = 0.
	\end{align}
The second derivatives of the potential are given by
	\begin{align}
	m_{\phi \phi}^{2} & \equiv \frac{\partial^{2}V}{\partial \phi^{2}} = \mu_{\phi}^{2} + 3\lambda_{\phi}\phi^{2} + \frac{\lambda_{\phi s}}{2} \sigma^{2}, \\
	m_{\sigma \sigma}^{2} & \equiv \frac{\partial^{2}V}{\partial \sigma^{2}} = \mu_{s}^{2} + 3\lambda_{s}\sigma^{2} + \frac{\lambda_{\phi s}}{2} \phi^{2}, \\
	m_{\phi \sigma}^{2} & \equiv \frac{\partial^{2}V}{\partial \phi \partial \sigma} = \lambda_{\phi s} \phi \sigma. 
	\end{align}
This gives a mass matrix in the $(\phi \; \sigma)$ basis
	\begin{align}
	\begin{pmatrix}
	m_{\phi \phi}^{2} & m_{\phi \sigma}^{2} \\
	m_{\phi \sigma}^{2} & m_{\sigma \sigma}^{2}
	\end{pmatrix},
	\end{align}
with eigenvalues
	\begin{equation}
	m_{\phi (\sigma)}^{2} = \frac{1}{2}\left(m_{\sigma \sigma}^{2}+m_{\phi \phi}^{2}+(-)\sqrt{(m_{\phi \phi}^{2}-m_{\sigma \sigma}^{2})^{2}+4m_{\phi \sigma}^{4}}\right).
	\label{eq:higgseigen}
	\end{equation}
At the stationary point $(v_{\phi}, v_{s})$ these are simply
	\begin{equation}
	m_{\phi (\sigma)}^{2} = \lambda_{\phi}v_{\phi}^{2}+\lambda_{s}v_{s}^{2}+(-)\sqrt{(\lambda_{\phi}v_{\phi}^{2}-\lambda_{s}v_{s}^{2})^{2}+(\lambda_{\phi s}v_{\phi}v_{s})^{2}},
	\end{equation}
where we will associate $m_{\phi}$ with the SM Higgs, $m_{\phi}=125$ GeV. We write the mixing between the fields $\phi$ and $\sigma$ into the physical states $\phi^{'}$ and $\sigma^{'}$ in the usual way
	\begin{equation}
	\begin{pmatrix} \phi^{'} \\ \sigma^{'} \end{pmatrix} = \begin{pmatrix}  \cos{\theta} & \sin{\theta} \\ -\sin{\theta} &  \cos{\theta} \end{pmatrix} \begin{pmatrix} \phi \\ \sigma \end{pmatrix},
	\end{equation}
where the mixing angle is given by
	\begin{equation}
	\tan{2\theta}=\frac{\lambda_{\phi s}v_{\phi}v_{s}}{\lambda_{\phi}v_{\phi}^{2}-\lambda_{s}v_{s}^{2}} \, .
	\end{equation}
The mixing angle will be negligibly small in our model, so we will not distinguish between $\phi$ and $\phi^{'}$ and $\sigma$ and $\sigma^{'}$ below. To ensure $(v_{\phi}, v_{s})$ is a minimum, we require a positive determinant of the mass matrix at this point. This is achieved for
\be
\label{eq:minimum}
\lambda_{\phi s}^{2} < 4\lambda_{s}\lambda_{\phi} \, .
\ee
Using the relation (\ref{eq:parrel}) between $\lambda_{s}$ and $\lambda_{\phi s}$, this translates into
	\begin{align}
	\label{eq:lphislim}
	\lambda_{\phi s} < \frac{2v_{\phi}^{2}\lambda_{\phi}}{C\Lambda_{s}^{2}} = 1.7\times10^{-20}\left(\frac{\lambda_{\phi}}{0.13}\right)\left(\frac{0.92}{C}\right)\left(\frac{10^{12} \; \mathrm{GeV}}{\Lambda_{s}}\right)^{2}, \\
	\lambda_{s} < \frac{v_{\phi}^{4}\lambda_{\phi}}{C^{2}\Lambda_{s}^{4}} = 5.6 \times 10^{-40} \left(\frac{\lambda_{\phi}}{0.13}\right)\left(\frac{0.92}{C}\right)^{2}\left(\frac{10^{12} \; \mathrm{GeV}}{\Lambda_{s}}\right)^{4}. \label{eq:lslim}
	\end{align}
According to (\ref{eq:lslim}) we are always in the regime in which 
\be
\lambda_{s}v_{s}^{2} \ll \lambda_{\phi}v_{\phi}^{2} \, 
\ee 
such that  $m_{\sigma}^{2} \simeq 2\lambda_{s}v_{s}^{2}$. Using this and eqs.~(\ref{eq:parrel}) and (\ref{eq:lphislim}), one finds
	\begin{equation}
	\label{eq:sigmamasslim}
	m_{\sigma} < \frac{2\sqrt{\lambda_{\phi}}\epsilon_{s}v_{\phi}^{2}}{C\Lambda_{s}} = 9 \; \mathrm{eV} \; \left(\frac{\lambda_{\phi}}{0.13}\right)^{1/2}\bigg(\frac{\epsilon_{s}}{0.2}\bigg)\left(\frac{0.92}{C}\right)
	\left(\frac{10^{12} \; \mathrm{GeV}}{\Lambda_{s}} \right) \, ,
	\end{equation}
which can also be written as 
\be
m_{\sigma}  \Lambda_s < \left(\frac{\lambda_{\phi}}{0.13}\right)^{1/2}\bigg(\frac{\epsilon_{s}}{0.2}\bigg)\left(\frac{0.92}{C}\right)    \times 10^{-2} \  {\mbox{TeV}}^2\, ,
\ee

This inequality is incompatible with (\ref{mainconstraint}) which follows from the meson oscillation constraints if $m_{\sigma} > m_K$.
Therefore, a weak scale flavon leading to Yukawa variation during the EW phase transition
seems incompatible with flavour constraints, assuming the simple scalar potential (\ref{eq:PotmodelA}).
An ultra light flavon ($m_{\sigma} < m_K$) is however compatible with flavour and cosmological constraints
as discussed in Appendix  \ref{section:flavorconstraints}, provided that $\Lambda_s \gtrsim 10^{12}$ GeV.
In this case, the bounds (\ref{eq:lphislim}) and (\ref{eq:lslim}) lead to extremely small quartic couplings. This essentially rules out this model as a reasonable illustration of 
naturally varying Yukawa couplings during the EW phase transition. Nevertheless, for the sake of illustration, we will proceed to study the phase transition for this tuned model to show how the desired cosmological evolution can be realized (notice that in Ref.~\cite{Berkooz:2004kx} the effect of varying Yukawas on the nature of the EW phase transition was not discussed and the quoted cosmological bound on $\Lambda_s$ was also different). 

We note that the root of the problem comes from equation (\ref{eq:minimum}) which  forces the flavon to be light to have an impact on Yukawa variation during the EW phase transition.
A way out would be to modify  the flavon potential (\ref{eq:PotmodelA}), for example if the running of $\lambda_s$ induces a minimum for $\sigma$ at a parametrically larger value, ${\cal O}$(TeV), than the EW scale. We do not pursue this here but come back to this point in our conclusion.

\subsection{The phase transition}
The computation of the high temperature effective scalar potential is described in Ref.~\cite{Baldes:2016rqn}.
We  include all quark species as strongly coupled fermions can qualitatively change the nature of the phase transition. The dependence on $\phi$ and $\sigma$ enters through the $\phi$ and $\sigma$ dependent masses of the fermions and Higgs bosons and the $\phi$ dependent $W^{\pm}$ and $Z$ boson masses. The fermion masses are of the form
	\begin{align}
	m_{fi}(\phi,\sigma)=\left|y_{i}\left(\frac{\sigma}{\sqrt{2}\Lambda_{s}}\right)^{n_{i}}\frac{\phi}{\sqrt{2}}\right|,
	\end{align}
where $n_{i}$ depends on the choice of Froggatt--Nielsen charges, $y_{i}$ is a dimensionless coupling. We assume the charge assignments above, i.e. 
	\begin{align}
	n_{i=t,b,c,s,u,d}&= \{0, \; 2, \; 3, \; 4, \; 7, \; 6\} \label{eq:ni},
	\end{align}
for the top, bottom, charm, strange, up and down quarks. For simplicity, we ignore the $\sigma$ and $\phi$ dependent mixing of the electroweak quark eigenstates.

We calculate the critical temperature, $T_{c}$, defined as the temperature at which the minima at $\phi=0$ and $\phi \neq 0$ are degenerate. The critical VEV $\phi_{c}$ is the value of $\phi$ at the second minimum at $T_{c}$. Successful EWBG requires a strong first-order phase transition, characterised by $\phi_{c}/T_{c} \gtrsim 1$. 
We first set all $y_{i\neq t}=1$. This results in a strong first-order phase transition with $\phi_{c}/T_{c}=1.5$. 
The large effective Yukawas increase the strength of the phase transition as extensively discussed in Ref.~\cite{Baldes:2016rqn}.

We next set the Yukawa couplings in the symmetric phase to
	\begin{align}
	y_{i=t,b,c,s,u,d}&= \{0.99, \; 0.60, \; 0.93, \; 0.34, \; 1.0, \; 0.43\}, \label{eq:yi} 
	\end{align}
which returns the observed fermion masses at $T=0$. We now find a crossover. This shows the sensitivity of the phase transition strength to the Yukawa couplings. In particular the strength of the phase transition is most sensitive to the bottom quark Yukawa. The reason is simply that the index $n_{b}=2$ is the smallest after  the top, this leads to fewer factors of $\sqrt{2}$ suppressing the effective Yukawa at $\sigma = \Lambda_{s}$ than for the other quarks.

The short summary of this analysis is that the strength of the phase transition 
mainly depends on what happens to the bottom quark yukawa and it 
increases with $y_{b}$ (we assume like in all FN models that the top quark yukawa is not controlled by the VEV of the FN field).

We have also tried modifying the potential so that $\sigma$ starts off at different values. In general, the larger the starting point for $\sigma$, the larger the effective Yukawas and the stronger the phase transition. Note that due to mixing effects, the Yukawa couplings do not necessarily have to take the values given in eq.~(\ref{eq:yi}), an $\mathcal{O}(1)$ cancellation could reduce the eigenvalue associated with the $b$ quark at $T=0$ compared to $T=100$ GeV (the indices $n_{i}$ should also be considered approximate as we are ignoring mixing effects). 

In conclusion, flavor constraints on this most minimal model have pushed us into a tuned  regime of parameter space where $\Lambda_{s} \gg v_{\phi}$ and $\lambda_{\phi s}, \; \lambda_{s} \ll 1$. 
The values  for the quartic couplings (\ref{eq:lphislim}) and (\ref{eq:lslim}) are clearly untenable. 
We kept this example as a pedagogical step and 
now go on to explore more natural possibilities.

\begin{figure}[t]
\begin{center}
\includegraphics[width=250pt]{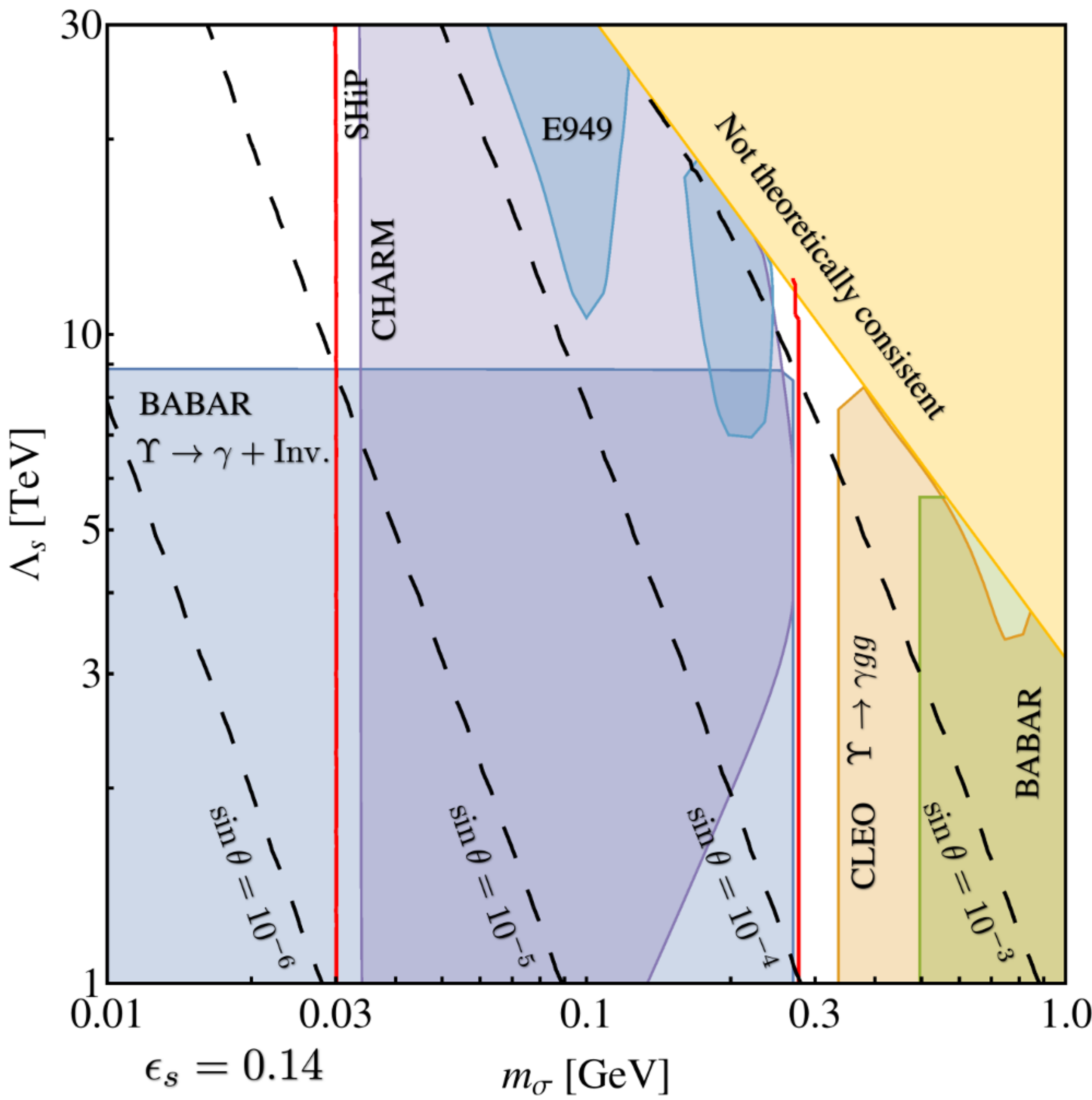}
\end{center}
\caption{\small Parameter space of the low scale scenario A-2 discussed in section~\ref{sec:lowscale} for $\epsilon_{s}=0.14$. The limits come from the CHARM beam dump experiment~\cite{Bergsma:1985qz,Bezrukov:2009yw,Clarke:2013aya}, searches for $\Upsilon \to \gamma + \mathrm{Inv.}$ at BABAR~\cite{delAmoSanchez:2010ac}, $\Upsilon \to \gamma gg$ at BABAR~\cite{Lees:2013vuj} and CLEO~\cite{Besson:2005jv,McKeen:2008gd} and limits on $K \to \pi + \mathrm{Inv.}$ from E949~\cite{Artamonov:2009sz,Clarke:2013aya}. We also show the projected sensitivity of SHiP~\cite{Alekhin:2015byh}. Limits from BBN~\cite{Krnjaic:2015mbs,Cadamuro:2011fd}, SN1987A~\cite{Krnjaic:2015mbs,Ishizuka:1989ts}, $\Upsilon \to \gamma \mu \mu$~\cite{Love:2008aa,McKeen:2008gd}, $B \to K + \mathrm{Inv.}$~\cite{Clarke:2013aya}, $B \to K \mu \mu$~\cite{Clarke:2013aya,Schmidt-Hoberg:2013hba}, $B \to K\pi\pi$~\cite{BABAR:2011aaa} and $K \to \pi \mu \mu$~\cite{Batley:2011zz,Clarke:2013aya} are too weak to appear on the plot. The enhanced coupling of $\sigma$ to $b$ quarks from eq.~(\ref{eq:enhancedb}) increases the sensitivity to $\Upsilon$ decays and also modifies induced couplings to $\gamma \gamma$, $gg$, pions and nucleons compared to the usual scenario of a light scalar mixing with the SM Higgs~\cite{Schmidt-Hoberg:2013hba,Clarke:2013aya,Krnjaic:2015mbs}.
}
\label{fig:unconstrained_1}
\end{figure}

\subsection{Scenario with disentangled hierarchy and mixing sectors}
\label{sec:lowscale}
Recently, a scenario was proposed, which allows one to disentangle the sector responsible for the hierarchy of masses and the sector responsible for the observed mixing angles~\cite{Knapen:2015hia}. The scenario is far from minimal as it introduces a separate `flavon' for each quark. 
In this construction, the SM fermions are not charged under some new horizontal symmetry, but the light flavons responsible for the mass hierarchy are. In the Yukawa coupling, the charge of the light flavon is canceled by the charge of some  other scalar part of the sector controlling the flavor violation. The aligned structure in this latter sector is such that, effectively, the flavon only couples to the mass eigenstates.
Assuming such a construction can be made, the limits on the mass hierarchy sector are then greatly relaxed as each flavon couples to the individual mass eigenstates only. Such a scenario is also amenable to Yukawa variation. The high dimensionality of the scalar potential on the other hand precludes a simple analysis of the phase transition. For simplicity, we consider only the variation of a single flavon and hence a single Yukawa coupling in this section. The scenario is then almost identical to the discussion of model A-1 but can occur for a much smaller  $\Lambda_{s}$ scale (hence a higher $m_{\sigma}$). The main advantage of such a scenario is that we can avoid the huge separation of scales and extremely small couplings required in the previous model A-1 example. However, we shall see that other experimental limits already severly constrain this scenario.

To illustrate the low scale mechanism, let us consider the variation of the bottom Yukawa only through the term
	\begin{equation}
	\label{eq:enhancedb}
	\mathcal{L} \supset \frac{y_{b}}{\sqrt{2}}\left( \frac{\sigma}{\sqrt{2}\Lambda_{s}} \right)^{2}\phi \overline{b}b,
	\end{equation}
where the bottom quarks are in the mass eigenstate basis. Though this scenario is no longer constrained by flavor physics, it unavoidably leads to a light flavon which mixes with the SM Higgs. It is therefore constrained by light scalars searches. The parameter space for $\epsilon_{s}=0.14$ is shown in figure~\ref{fig:unconstrained_1}. 

Given that the value of $\epsilon_{s}$, defined in (\ref{eq:epsdefined}), is required to explain the bottom quark mass only and not the CKM mixing angles, we have greater freedom in its choice. We will consider different possibilities for $\epsilon_{s}$ and hence $y_{b}$. 
Like in model A-1, once the scale $\Lambda_{s}$, the final VEV $v_{\sigma}$ (or equivalently $\epsilon_{s}$) is set, the need for Yukawa variation is taken into account and a mass $m_{\sigma}$ is chosen, we can then solve for $\lambda_{s}, \lambda_{s \phi}$ and $\sin{\theta}$. Given $\Lambda_{s}$ and $v_{\sigma}$ there is a maximum $m_{\sigma}$ for which a consistent solution is possible. Below this two possibilities for $\lambda_{s}$ (and hence $\lambda_{s \phi}$ and $\sin{\theta}$) exist. In our analysis we take the smaller choice. The larger choice corresponds to values close to the stability bound~(\ref{eq:lslim}) and hence represents only a small, fine tuned, area of parameter space. The area in which a given $m_{\sigma}$ cannot be consistently achieved is also indicated in figure~\ref{fig:unconstrained_1}.

\begin{figure}[t]
\begin{center}
\includegraphics[width=200pt]{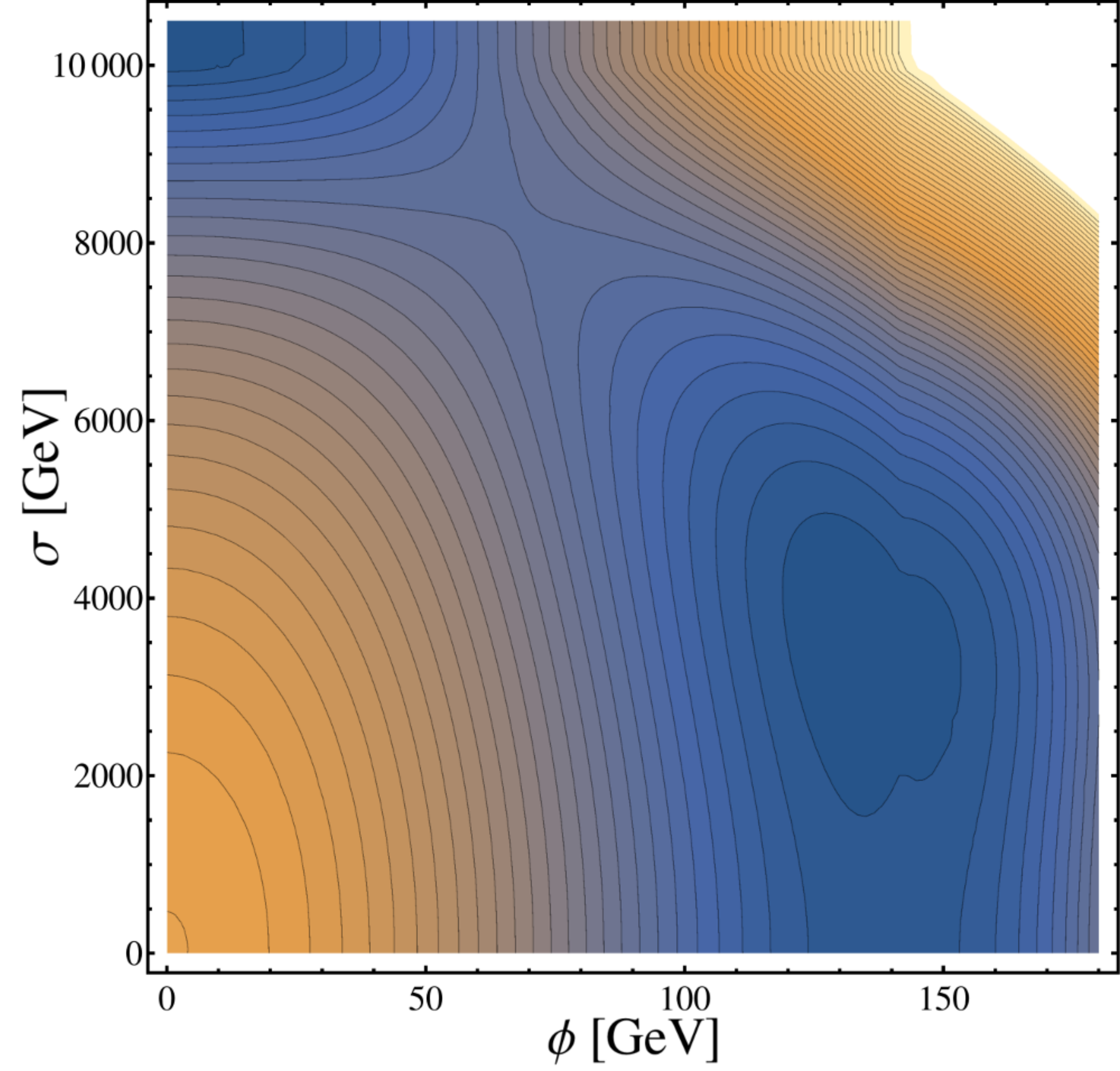}
\includegraphics[width=200pt]{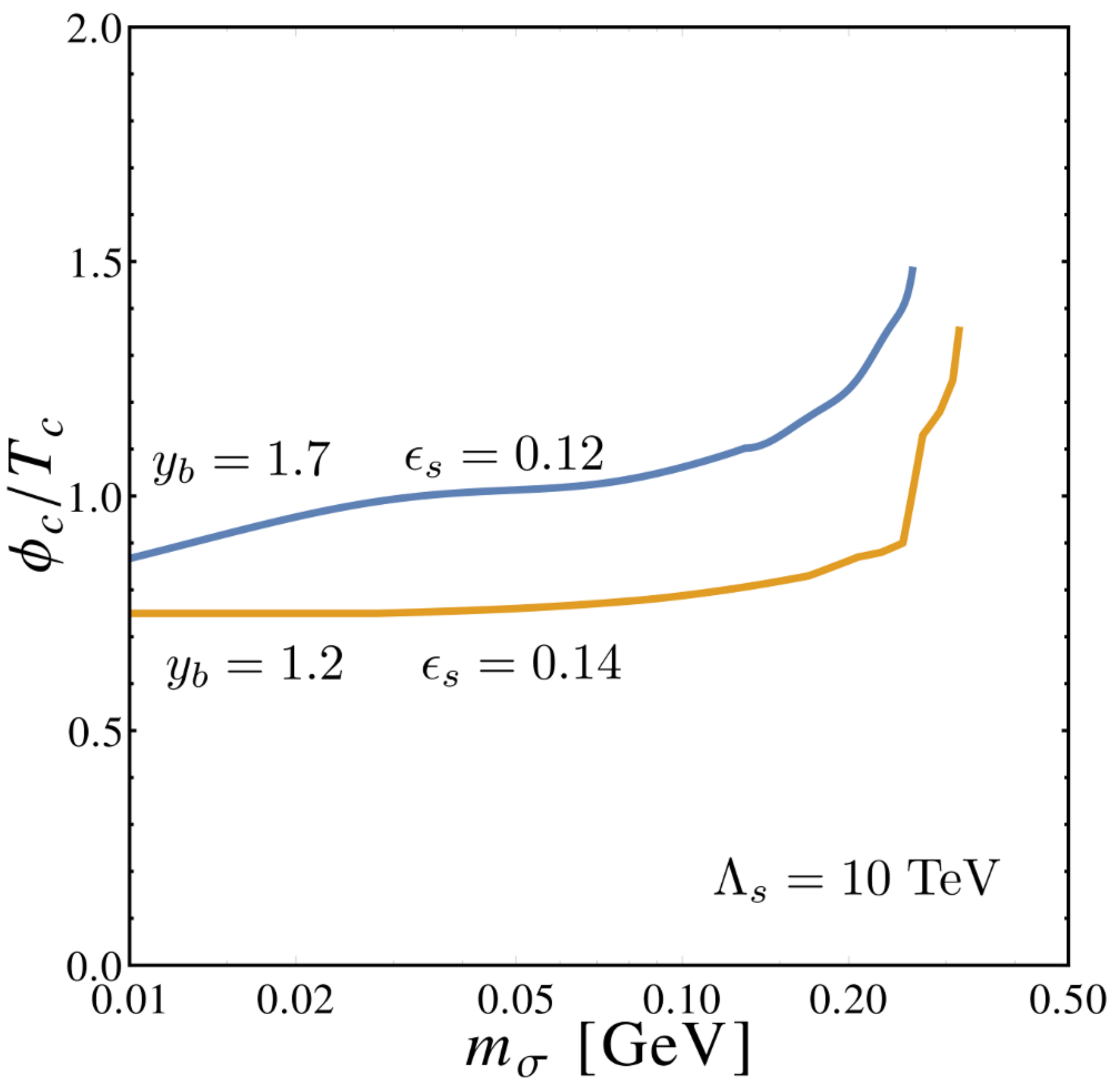}
\end{center}
\caption{\small Left: The effective potential at the critical temperature, in the low scale scenario discussed in section~\ref{sec:lowscale}, for $\Lambda_{s}=10$ TeV, $m_{\sigma}=0.03$ GeV, $\epsilon_{s}=0.12$, $y_{b}=1.7$, $\lambda_{\phi s}=10^{-6.3}$, $ \lambda_{s}=1.6 \times10^{-10}$. We find a strong first-order phase transition with $\phi_{c}\simeq T_{c} =144$ GeV. Right: Strength of the phase transition for different choices of $\epsilon_{s}$ (and hence $y_{b}$). As expected larger values of $y_{b}$ lead to stronger phase transitions. Note the area between $0.03$ GeV $\lesssim m_{\sigma} \lesssim 0.3$ GeV is ruled out by CHARM.}
\label{fig:unconstrained_2}
\end{figure}

The coupling $y_{b}$ is set by our choice of $\epsilon_{s}$ and the observed bottom quark mass. Different choices of $\epsilon_{s}$ can lead to different phase transition strengths. This is shown in figure~\ref{fig:unconstrained_2}. We see that phase transitions with $\phi_{c}/T_{c} \approx 0.7$ $(1.0)$ can be achieved with $\epsilon_{s}=0.14$ $(0.12)$. Note that due to possible stability issues with the potential for large Yukawa couplings, we have assumed the effective Yukawa coupling saturates at $\sigma = \Lambda_{s}$ and  we only search for minima in the region $0 \leq \sigma \leq 1.2 \Lambda_{s}$ and $0 \leq \phi \leq 300$ GeV. Outside this region we assume higher dimensional operators stabilise the potential. (For $\epsilon_{s}=0.12$, the instability scale at $T=0$ is safely at $\phi = 1200$ GeV along $\sigma=\Lambda_{s}$. But for $\epsilon_{s}=0.11$, the $T=0$ one loop potential runs away to infinity along the $\phi$ direction at $\sigma=\Lambda_{s}$, with no barrier present. Hence we do not consider smaller values for $\epsilon_{s}$.)

The summary of this analysis is that only a very light flavon can escape all constraints:
\be
m_{\sigma} \lesssim 1 \mbox{ GeV and } \Lambda_s \lesssim {\cal O} (10 \mbox{ TeV}) \, .
\ee
We cannot set $\epsilon_s$ smaller than 0.12 as this would require too large $y_b$ to obtain the correct bottom quark mass with a stable potential. On the one hand, the instability at large field values in the scalar potential scales as $y_b^4$. 
On the other hand, increasing $\Lambda_s$ (as required by experimental constraints) pushes us to smaller $\lambda_{s}$, $\lambda_{s \phi}$ and $m_{\sigma}$. 
From Eq.~(\ref{eq:enhancedb}), radiative corrections to the scalar mass from a $b$ quark loop integrated to the FN scale are expected to be $\Delta m_{\sigma}^{2} \sim (m_{b}/(4\pi\epsilon_{s}))^{2}$, hence the small masses required to be consistent with experimental constraints are in a tuned regime.

Note that to obtain a massive pseudo-Nambu-Goldstone Boson we have explicitly broken $U(1)_{\rm FN}$ as discussed in section~\ref{sec:FNexplained}. The remnant $Z_{2}$ should now also be explicitly broken in order to avoid domain walls. As this scenario is already highly constrained, we do not pursue this further here, but note that the issue is absent in models B which we shall discuss in Section \ref{sec:modelsB}.


\subsubsection{Exotic Higgs decays}
\label{sec:exohigssa}
In this model, the Higgs can decay into the flavon with the partial width
	\begin{equation}
	\Gamma(\phi \to \sigma \sigma)=\frac{\lambda_{\phi s}^{2} v_{\phi}^{2}}{32\pi m_{\phi}}\sqrt{1 - \frac{4m_{\sigma}^{2}}{m_{\phi}^{2}} }.
	\end{equation}
From the combination of ATLAS and CMS data, the Higgs signal yield is $\mu = 1.09 \pm 0.11$~\cite{ATLAS-CONF-2015-044}. In terms of the SM Higgs width, this limits the total width to $\Gamma_{\phi} \lesssim 1.15 \; \Gamma_{\phi}^{\rm SM}$, assuming no new contributions to the production cross section. In the parameter space of interest to us here, we may ignore the final state masses.  We then find a limit on the portal coupling $\lambda_{\phi s} \lesssim 0.011$. Other experimental results are currently more constraining (in the allowed areas of figure~\ref{fig:unconstrained_1}, we find $\lambda_{\phi s} \lesssim 10^{-3}$.)

Another possible decay is  $\phi \to \overline{b}b\sigma$ with the width
        \begin{equation}
        \Gamma(\phi \to \overline{b}b\sigma) = \frac{\epsilon_{s}^{2}|y_{b}|^{2}}{256\pi^{2}}\frac{m_{\phi}^{3}}{\Lambda_{s}^{2}} \, .
        \end{equation}
Using the relation between $m_{b}$, $y_{b}$ and $\epsilon_{s}$, this implies the branching fraction         \begin{equation}
        \mathrm{Br}(\phi \to \overline{b}b\sigma) = 1.1 \% \left( \frac{0.1}{\epsilon_{s}} \right)^{2} \left( \frac{1 \; \mathrm{TeV} }{\Lambda_{s}} \right)^{2}.
        \end{equation}
        Such decay channel is weakly constrained.

In conclusion, Model A-2 is a testable possibility realizing varying Yukawas at the EW scale. It relies on the specific mechanism recently presented in Ref.~\cite{Knapen:2015hia} in which EW scale flavons  do not lead to low energy flavor violating effects. However, as is clear from figure~\ref{fig:unconstrained_1}, the example we have considered here is severely constrained by other experimental searches and
 the required light flavon mass is not stable under radiative corrections.
This construction has not yet been fully explored and it would be interesting to see alternative complete explicit implementations.

\section{Models B: Two Froggatt--Nielsen field models}
\label{sec:modelsB}

Having presented the challenges associated with single flavon models where the flavon VEV varies during the EW phase transition,
we now move on to a model with two Froggatt--Nielsen scalars,  carrying either different or the same FN charges. The first scalar $S$ generates the Yukawa hierarchy today while the second one $X$ has a negligible VEV at low temperatures but develops a VEV at early times and plays a role during the EW phase transition.  As we will see, this is quite a generic situation. The main advantage of such a scenario is that we can avoid the huge separation of scales and extremely small scalar quartic couplings required in the previous model A-1 without the need for disentangling the Yukawa hierarchy and mixing sectors like in A-2. 
Equation (\ref{eq:FNyuks1}) now becomes
\begin{align}
	\label{eq:FNyuks1B}
	\mathcal{L} & = \tilde{y_{ij}}\left(\frac{S }{ \Lambda_{s} }\right)^{\tilde{n}_{ij}} \overline{Q}_i \tilde{\Phi} U_{j} + y_{ij}\left(\frac{S }{ \Lambda_{s} }\right)^{n_{ij}} \overline{Q}_i \Phi D_{j} 	 \nonumber \\
		& \; + \tilde{f_{ij}}\left(\frac{X }{ \Lambda_{\chi} }\right)^{\tilde{m}_{ij}} \overline{Q}_i \tilde{\Phi} U_{j} + f_{ij}\left(\frac{ X }{ \Lambda_{\chi} }\right)^{m_{ij}} \overline{Q}_i \Phi D_{j} + H.c.,
	\end{align}  
where $f_{ij}$ and  $\tilde{f_{ij}}$  are dimensionless couplings, also assumed to be $\mathcal{O}(1)$ while 
 $m_{ij}$ and $\tilde{m}_{ij}$  are chosen to form singlets under $U(1)_{\rm FN}$.
We assume either $\langle X \rangle = 0$ or $\langle X \rangle \ll \Lambda_{\chi}$ today, but we will consider cosmological histories with non-negligible $\langle X \rangle$ below.
We also define
\begin{equation} 
	\label{eq:epsdefinedB}
	 \qquad \epsilon_{\chi} \equiv \frac{ \langle X \rangle }{ \Lambda_{\chi} }=\frac{v_{\chi}}{\sqrt{2}\Lambda_{\chi}},
	\end{equation}
and  the new scalar field reads
	\begin{equation}
	\label{eq:recomponentsB}
	 \qquad X = \frac{\chi + i \eta}{\sqrt{2}} \, .
	\end{equation}
In this setup, the scalar $S$ no longer has a large impact on the nature of the EWPT nor does it induce large CP-violating sources, as its VEV  does not vary significantly during the EWPT. It is therefore just a spectator during the  EWPT. In this case, there is no longer any tension with flavor constraints and we expect 
$m_{\sigma} \sim 1 $ TeV.
The main cosmological actor whose VEV is varying during the EWPT is instead $\chi$ which is  therefore expected to be light, below the EW scale. Since $\chi$ has a negligible VEV today, this is no longer in conflict  with flavor constraints.  

The details of this mechanism depend on the FN charge of $X$. If $X$ and $S$ carry the same charge (model B-2), the model appears to be more minimal but it is also more constrained by meson oscillations, as discussed  in Appendix \ref{sec:constraints2}. If $Q_{FN}(\chi)=-1/2$ (model B-1) then meson oscillation constraints are considerably relaxed (from $\Lambda_{\chi} \gtrsim 2$ TeV to $\Lambda_{\chi}\gtrsim 700$ GeV) but the model features twice as many FN fermions which are also constrained by the LHC.

\subsection{Models B-1: $Q_{FN}(\chi)=-1/2$ }

\subsubsection{Tree level potential} 

Consider the scalar potential 
	\begin{align}
\label{eq:2FieldPot}
	V & = \mu_{\phi}^{2}\Phi^{\dagger}\Phi+\lambda_{\phi}(\Phi^{\dagger}\Phi)^{2} + \mu_{s}^{2}S^{\dagger}S+\lambda_{s}(S^{\dagger}S)^{2} + \mu_{\chi}^{2}X^{\dagger}X+\lambda_{\chi}(X^{\dagger}X)^{2}  \nonumber \\
	   & + \lambda_{\phi s}(\Phi^{\dagger}\Phi)(S^{\dagger}S) + \lambda_{\phi \chi}(\Phi^{\dagger}\Phi)(X^{\dagger}X) + \lambda_{\chi s}(X^{\dagger}X)(S^{\dagger}S) \\
	   & + \tilde \mu_{\chi s} (X^{\dagger}X^{\dagger}S + H.c.). \nonumber
	\end{align}
For simplicity we assume negligible mixing between $S$ and $\Phi$ and also assume only negligible changes of $v_{s}$ during the EWPT. We also only consider renormalisable interactions, which are sufficient to illustrate our mechanism. We explain in Appendix \ref{sec:dimension6} how corrections to the potential from dimension-6 operators do not qualitatively  change our conclusions. Note the $\tilde \mu_{\chi s}$ term together with non-zero $v_{s}$ results in a mass splitting between the real and imaginary components of $X$ (we could also include an explicit breaking but it is superfluous here). Again we ignore the effect of the pseudoscalar, which is anyway expected to have only negligible effects on the effective potential. 

It should be noted that a choice of FN charge $-1$ for $X$ would allow for terms such as $S^{\dagger}S^{\dagger}SX$ in the Lagrangian, resulting in a too large VEV for $\chi$ for the mechanism to be consistent with flavor constraints, unless such terms are tuned away. The choice of FN charge $-1/2$ for $X$ ensures a $Z_{2}$ symmetry is present both in eq.~(\ref{eq:2FieldPot}) and in the effective Yukawa sector ($X$ being odd and all other fields even). Later we shall see that a scenario with an exact $Z_{2}$ symmetry and $\langle \chi \rangle = 0$ is highly constrained by its dark matter phenomenology. We shall therefore consider explicit breaking of the $Z_2$ below; as a result $\chi$ will gain a small VEV and become unstable. However, for purposes of illustration it is simpler to consider the exact $Z_2$ case first, and only later introduce a small breaking, once the phenomenological issues have been identified.

Starting from (\ref{eq:2FieldPot}), we redefine the $\mu_{\phi}$ and $\mu_{\chi}$ parameters to absorb the contribution from $v_{s}$ to the quadratic $\phi$ and $\chi$ terms. The relevant part of the scalar potential for our analysis of the phase transition then becomes
	\begin{align}
	\label{eq:potsimple}
	V  & = \frac{ \mu_{\phi}^{2} }{ 2 }\phi^{2}+ \frac{ \lambda_{\phi} }{ 4 } \phi^{4} + \frac{ \mu_{\chi}^{2} }{ 2} \chi^{2}+\frac{ \lambda_{\chi} }{ 4 }\chi^{4} + \frac{ \lambda_{\phi \chi } }{ 4 } \phi^{2}\chi^{2}.
	\end{align}
For $\phi=0$ we require a minimum at $\chi = v_{\chi} \neq 0$, while for $\phi=v_{\phi}$ we require a minimum at $\chi=0$.\footnote{Note that $v_{\chi}$ denotes the tree level minimum in the $\chi$ direction at $\phi=0$, not the $T=0$ minimum.} The VEV conditions are
	\begin{align}
	\mu_{\chi}^{2}+\lambda_{\chi}v_{\chi}^{2}=0, \\
	\mu_{\phi}^{2}+\lambda_{\phi}v_{\phi}^{2}=0.
	\end{align}
We also require $\mu_{\chi}^{2}, \; \mu_{\phi}^{2} < 0$ and $\mu_{\chi}^{2} + \lambda_{\phi \chi } v_{\phi}^{2} > 0$. Thermal effects will make  $(\phi, \chi) \approx (0,v_{\chi})$ the global minimum of the potential at temperatures above the EWPT. The global minimum at zero temperature should be $(\phi,\chi)=(v_{\phi},0)$. This imposes the constraint
	\begin{equation}
	\label{eq:lxlim}
	\lambda_{\chi} < \lambda_{\phi} \left(\frac{v_{\phi} }{ v_{\chi} }\right)^{4} = 4.7\times10^{-4} \left(\frac{1 \; \text{TeV} }{ v_{\chi} } \right)^{4}.
	\end{equation}
This constraint is modified by one-loop effects, which we take into account when scanning over the parameter space of the model below. The mass of $\chi$ at the global minimum is given by
	\begin{equation}
	m_{\chi}^{2} = \mu_{\chi}^{2}+\frac{\lambda_{\phi \chi}v_{\phi}^{2}}{2}=-\lambda_{ \chi}v_{\chi}^{2}+\frac{\lambda_{\phi \chi}v_{\phi}^{2}}{2},
	\label{eq:modelbmass}
	\end{equation}
from which we see choices of $\lambda_{\phi \chi} > 2 \lambda_{ \chi}v_{\chi}^{2}/v_{\phi}^{2}$ are required to make this mechanism viable. An important remark here is that $m_{\chi}$ cannot be higher than the EW scale.
This is a generic prediction of models of varying Yukawas in the Froggatt--Nielsen context, as we further discuss in Appendix \ref{sec:dimension6}. 

\begin{figure}[t]
\begin{center}
\includegraphics[width=200pt]{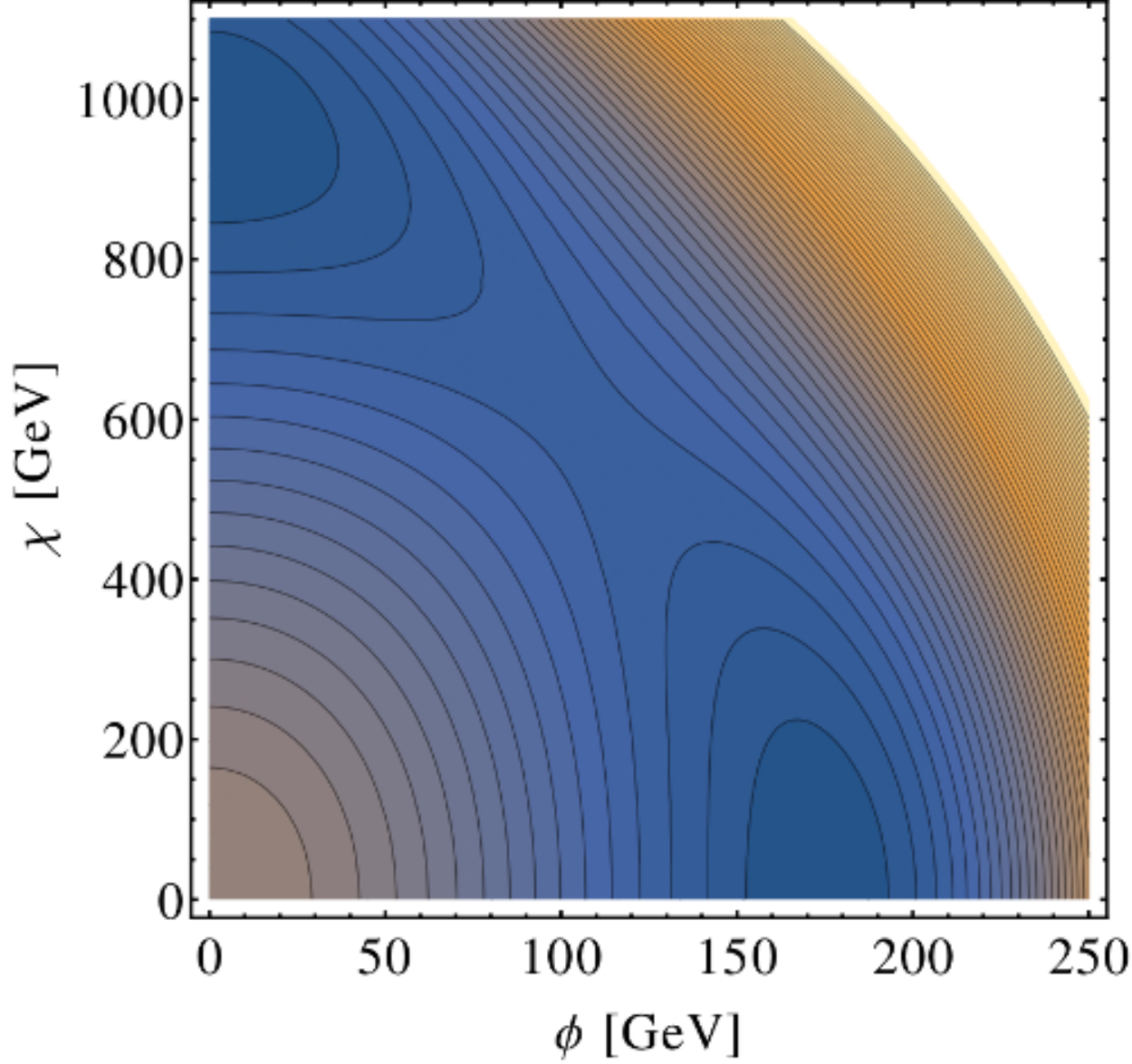}
\includegraphics[width=200pt]{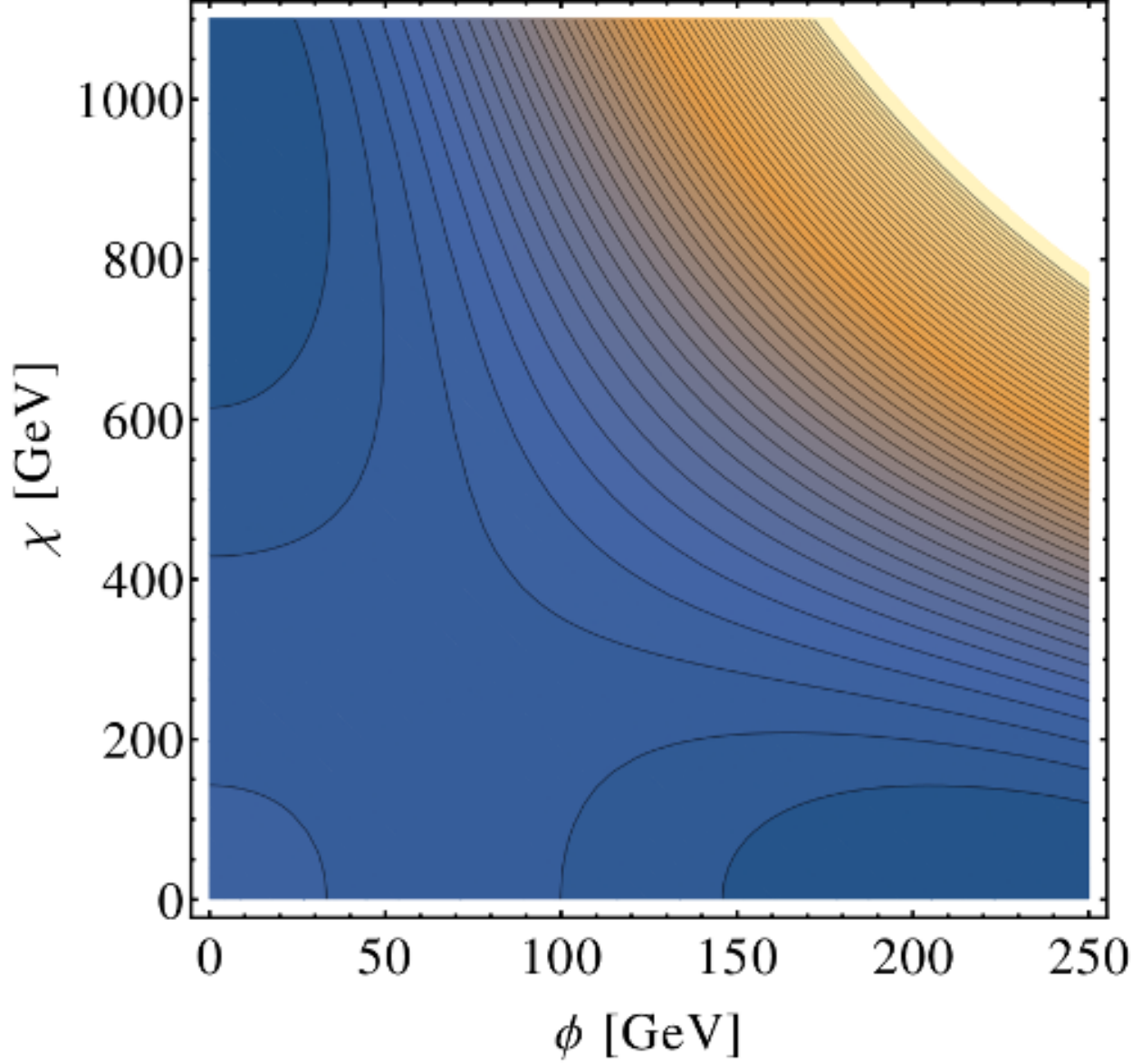}
\end{center}
\caption{\small Left: the potential at the critical temperature for $v_{\chi}=\Lambda_{\chi}=1$ TeV, $\lambda_{\chi}=10^{-4}$ and $\lambda_{\phi \chi}=10^{-2}$ ($m_{\chi}=14$ GeV) displaying two minima. We find $\phi_{c}=174$ GeV and $T_{c}=133$ GeV giving a strong first-order phase transition with $\phi_{c}/T_{c}=1.3$.  Right: same as left but with $\lambda_{\chi}=10^{-3.4}$, $\lambda_{\phi \chi}=10^{-0.8}$ ($m_{\chi}=137$ GeV), which yields $\phi_{c}=235$ GeV, $T_{c}=79$ GeV and $\phi_{c}/T_{c}=3.0$. Note the saddle point between the two minima generally moves towards the origin as we increase $\lambda_{\phi \chi}$.}
\label{fig:pottcri}
\end{figure}
\begin{figure}[t]
\begin{center}
\includegraphics[width=250pt]{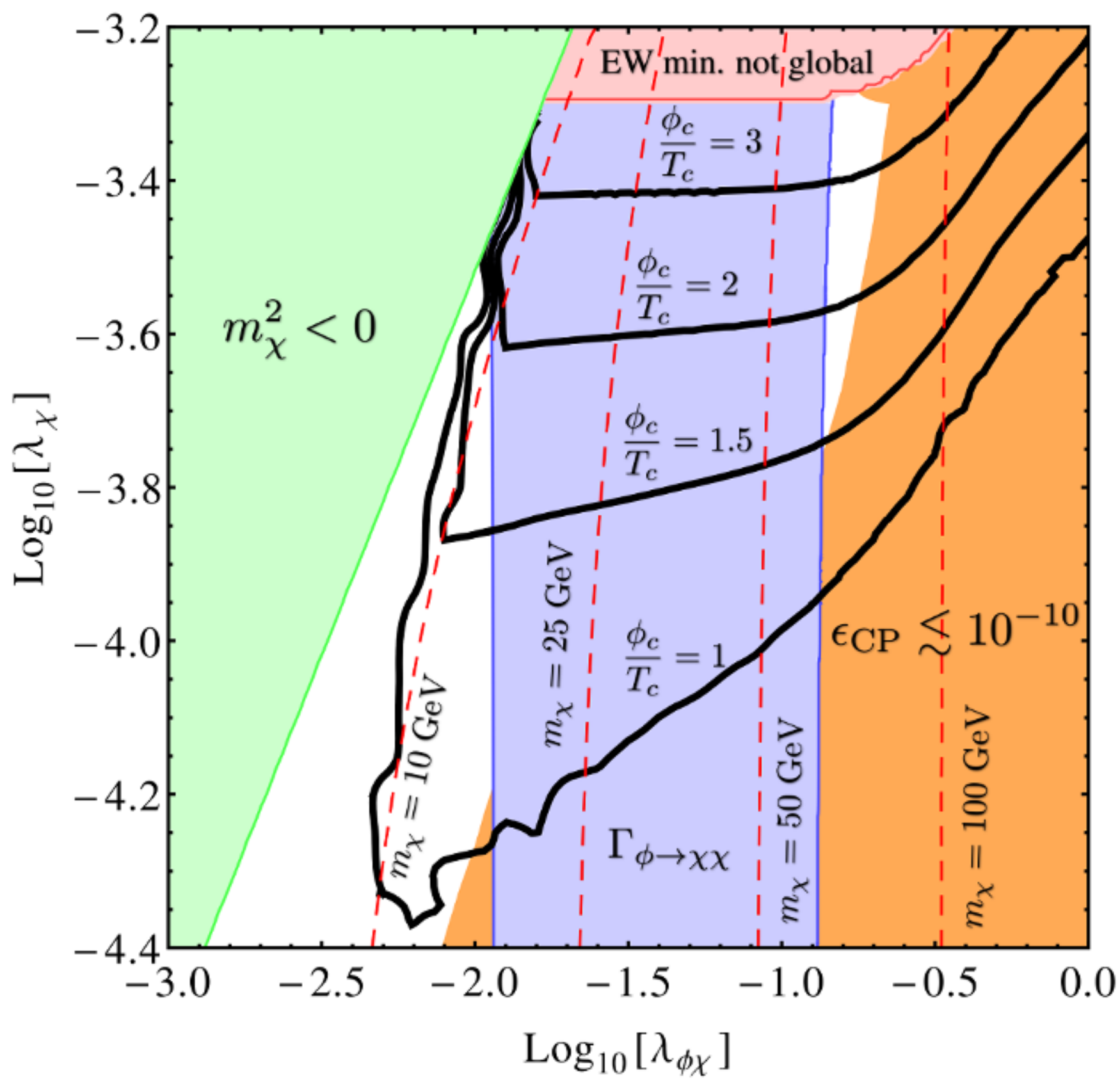}
\end{center}
\caption{\small Parameter space of the model for $v_{\chi}=\Lambda_{\chi}=1$ TeV and $f_{i \neq t}=1$. Black lines are contours of, from top to bottom, $\phi_{c}/T_{c}=3$, 2, 1.5, 1. Colored areas are excluded. The orange area indicates a phase transition starting from $\chi \lesssim 0.94\Lambda_{\chi}$, for which there may be insufficient CP violation, $\epsilon_{\rm CP} \lesssim 10^{-10}$ (using the naive estimate from the Jarlskog invariant). In the red and green shaded regions, the EW vacuum is not the global minimum (for the red shaded region there is a deeper minimum than the EW one along the $\chi$ axis, for the green shaded area, $m_{\chi}^{2}<0$ at the EW minimum, indicating there is a deeper minimum away from both the $\phi$ and $\chi$ axes). At points close to this limit, the phase transition occurs to a minimum away from the $\phi$ axis, which leads to drastically weaker phase transitions for $m_{\chi} \lesssim 10$ GeV. In the blue shaded area, the signal yield of the Higgs at the LHC is changed due to $\phi \to \chi \chi$ decays (see section~\ref{sec:exohiggs}). White area to the left is allowed due to the small coupling. 
The area to the right is allowed because the decay is kinematically disallowed. 
}
\label{fig:scan}
\end{figure}
\begin{figure}[t]
\begin{center}
\includegraphics[width=250pt]{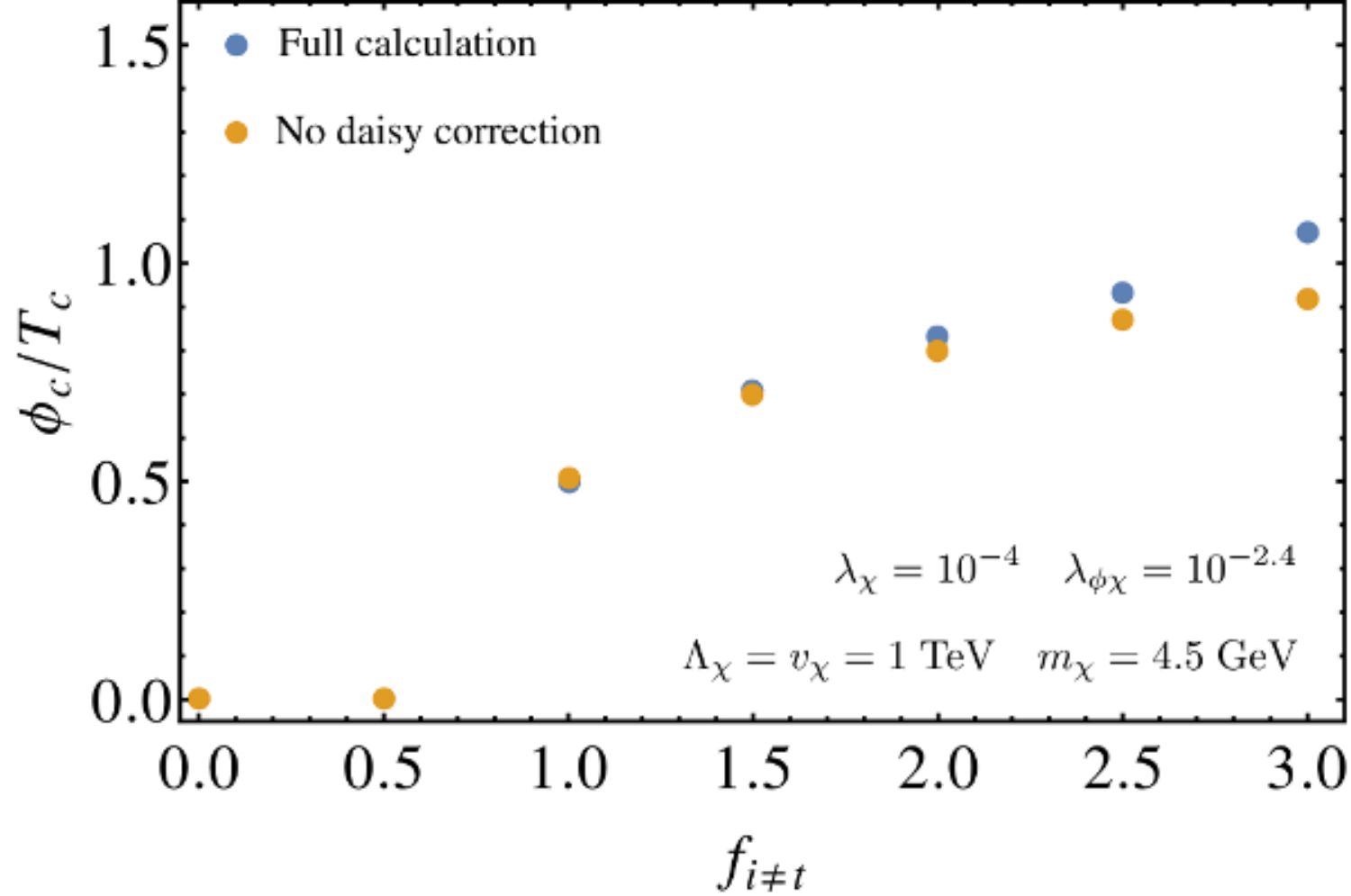}
\end{center}
\caption{\small Strength of the phase transition as a function of the Yukawa couplings for a choice of parameters. The calculation is repeated without the daisy correction, showing the daisy correction plays a subdominant role here in strengthening the phase transition (see related discussion in \cite{Baldes:2016rqn}).}
\label{fig:modb_yuk}
\end{figure}

\subsubsection{The phase transition }
We fix $\Lambda_{\chi}=1$ TeV throughout this section and explore the strength of the phase transition for different choices of $\lambda_{\chi}$ and $\lambda_{\phi \chi}$, for now we fix $f_{i \neq t}=1$. Examples of the potential at the critical temperature are shown in figure~\ref{fig:pottcri}. A scan over the parameter space, showing regions returning a strong first-order phase transition, is shown in figure~\ref{fig:scan}.

We now describe the dependence of the phase transition strength on the various parameters. First we note that the phase transition becomes stronger for $\lambda_{\chi}$ closer to the limit of eq.~(\ref{eq:lxlim}), as the two $T=0$ minima approach degeneracy. For values
	\begin{equation}
	\lambda_{\phi \chi} \geq -2 \frac{\mu_{\phi}^{2} }{ v_{\chi}^{2} } = 1.56 \times 10^{-2} \; \left( \frac{ \mathrm{TeV} }{ v_{\chi} } \right)^{2} 
	\end{equation}
the determinant of the mass matrix at $(\phi,\chi)=(0,v_{\chi})$ is positive and there is a tree level barrier between the two minima. Thermal effects can also lead to a strong first-order phase transition for smaller values of $\lambda_{\phi \chi}$. As we increase $\lambda_{\phi \chi}$ for fixed $\lambda_{\chi}$, the saddle point of the potential --- the minimum barrier height between the two minima --- moves toward the origin and the strength of the phase transition weakens. 

Large values of $\lambda_{\phi \chi}$ also push the location of the minimum on the $\chi$ axis toward the origin. The reason is the contribution of $\lambda_{\phi \chi} \chi^{2}/2$ to the Higgs mass and hence to the $T=0$ one-loop potential (the renormalization conditions used for the $T=0$ one-loop effective potential preserve the position of the EW minimum but not the position of the tree-level $(\phi,\chi)=(0,v_{\chi})$ minimum~\cite{Carena:2004ha}). Starting from smaller values of $\chi$ also means a suppression of CP violation. Note the $T=0$ one-loop effects also modify the bound from eq.~(\ref{eq:lxlim}). We identify two regions of parameter space  shown in figure~\ref{fig:scan}, one corresponds to $m_{\chi}\sim 60-70$ GeV and the other one correspond to $m_{\chi}$ between a few GeV and 20 GeV. So we typically deal with a light scalar $\chi$.

As we discussed previously, fermions with decreasing Yukawa couplings can also create thermal barriers, but in contrast with Model A, this is happening only in the restricted region of parameter space corresponding to small quartic couplings of $\chi$. We show the effect  on the strength of the phase transition in figure~\ref{fig:modb_yuk}. As expected, the phase transition becomes stronger as we increase the size of the dimensionless couplings $f_{i}$. The strongest effect comes again from the $b$ quark, as its effective Yukawa carries the lowest index, $n=4$, and hence is the least suppressed as $\chi$ decreases. Overall, the effect of the Yukawas is suppressed because of our choice of FN charge $-1/2$ for $X$, compared with the case with charge $-1$. The choice of charge which gives us the accidental $Z_{2}$ symmetry also doubles all the indices controlling the Yukawa couplings compared with the case for charge $-1$ and hence suppresses the effect of the Yukawas.

We now move to discuss phenomenological implications of this model.
These are linked to the light scalar state $\chi$. The other flavon is heavier, at the TeV scale and will be more difficult to probe. On the other hand, FN models also display new vector-like fermions which would appear here at the TeV scale and can be searched for. Note that in the case where both flavons have the same FN charge (model B-2), there will be one set of new fermions and $\Lambda_s \sim \Lambda_{\chi} \sim 1 $ TeV. While in the case where they have different charges (model B-1), there will be a doubling of the fermion states,  so this model is less minimal.

\subsubsection{Exotic Higgs decay $\phi \to \chi \chi$}
\label{sec:exohiggs}
In most of the allowed parameter space, the Higgs can decay into a $\chi \chi$.
The partial width is given by
	\begin{equation}
	\Gamma(\phi \to \chi \chi)=\frac{\lambda_{\phi \chi}^{2} v_{\phi}^{2}}{32\pi m_{\phi}}\sqrt{1 - \frac{4m_{\chi}^{2}}{m_{\phi}^{2}} }.
	\end{equation}
As for the Higgs decay discussed in Section~\ref{sec:exohigssa}, we apply the constraint $\Gamma_{\phi} \lesssim 1.15 \; \Gamma_{\phi}^{\rm SM}$, which we show in figure~\ref{fig:scan}. This limit is independent of the decay mode of $\chi$ (even if $\chi$ is stable on collider scales this limit is stronger than the direct constraint on invisible Higgs decays from vector boson fusion production~\cite{Aad:2015txa}).

\subsubsection{Explicit $Z_{2}$ breaking}
We have so far discussed the $Z_{2}$ symmetric case. However, if the $Z_2$ is exact then $\chi$ is stable and it cannot satisfy dark matter constraints
as shown in Appendix \ref{sec:darkmatterconstraints}. We have therefore 
 to consider explicit $Z_{2}$ breaking for $\chi$. For instance, we take the $T=0$ potential for $\chi$
	\begin{equation}
	V = a^{3}\chi + \frac{m_{\chi}^{2}}{2} \chi^{2} + \frac{b}{3}\chi^{3}+\frac{\lambda_{\chi}}{4}\chi^{4},
	\end{equation}
where nonzero $a$ or $b$ mean the $Z_{2}$ symmetry is broken. We wish to introduce a small breaking, in order not to run into conflict with flavor constraints. A non-zero $b$ typically results in a minimum deeper than the EW one at approximately
	\begin{equation}
	\langle \chi \rangle \approx \frac{-b \pm \sqrt{b^{2}-4\lambda_{\chi}\mu^{2}}}{2\lambda_{\chi}}.
	\end{equation}
Hence if $|b| \gtrsim 2\sqrt{\lambda_{\chi}}m_{\chi} \sim 0.1$ GeV, we have a deep minimum at $\langle \chi \rangle \gtrsim \frac{m_{\chi}}{\sqrt{\lambda_{\chi}}} \sim 1$ TeV. This VEV is both too large to be consistent with flavor constraints and also outside the domain of validity of our effective field theory. We therefore set $b \lesssim 2\sqrt{\lambda_{\chi}}m_{\chi}$ and consider non-zero $a$ instead. A small $a$ results in a minimum at
	\begin{equation}
	\langle \chi \rangle \approx -\frac{a^{3}}{m_{\chi}^{2}} \equiv v_{\chi a}.
	\end{equation}
(We include a subscript $a$ to distinguish this small $T=0$ VEV with the large VEV $v_{\chi}$ present before the EWPT). Once $\chi$ gains a VEV it mixes with the SM-like Higgs with mixing angle
	\begin{equation}
	\label{eq:modbmix}
	\theta \approx \frac{\lambda_{\phi \chi} v_{\chi a} }{ \lambda_{\phi}v_{\phi}} \approx  3\times 10^{-2} \times \lambda_{\phi \chi}  \bigg( \frac{ v_{\chi a} }{ 1 \; \text{GeV} } \bigg). 
	\end{equation}
The non-zero VEV for $\chi$ introduces new contributions at tree and loop level to the Wilson operators resulting in meson oscillations. We have checked that for a scale $\Lambda_{\chi} \sim 1$ TeV, VEVs as high as $v_{\chi a} \sim 100$ GeV for $m_{\chi}\sim 10$ GeV are not in conflict with the constraints from the UTfit colllaboration~\cite{Bona:2007vi,Derkach}.

\subsubsection{Decays of $\chi$ }
\label{sec:chidecaychargehalf}

The nonzero $v_{\chi a}$ allows $\chi$ to decay. This can happen either through its mixing with the Higgs, or directly through its Yukawa type interactions. Taking into account that the decay into $t\overline{c}$ is kinematically forbidden, we find the leading decay rate is $\chi \to \overline{c}c$. The leading contribution arises from the coupling
	\begin{equation}
	\mathcal{L} = \tilde{f}_{32} \left( \frac{X} {\Lambda_{\chi}} \right)^{2} \Phi \overline{c_{L}'}t_{R}' +H.c. \to \mathrm{Re}[U_{u}^{32 \ast}W_{u}^{22}\tilde{f}_{32}]\frac{ v_{\chi a}v_{\phi} }{ \sqrt{2}\Lambda_{\chi}^{2} } \chi \overline{c}c,
	\end{equation}
where we have explicitly denoted the flavor basis with primes and $U_{u}^{32}$ and $W_{u}^{22}$ are entries of the rotation matrices which bring the up-type quarks into the mass basis (see Appendix~\ref{sec:flavoncouplings}). Numerically we find $W_{u}^{22} \sim 1$ and $U_{u}^{32} \sim \epsilon_{s}^{2}$. Hence the above coupling is of the order
	\begin{equation}
	\mathcal{L} \sim  10^{-6}\tilde{f}_{32} \left( \frac{v_{\chi a} }{ 1 \; \mathrm{GeV} } \right) \left( \frac{ 1 \; \mathrm{TeV} }{ \Lambda_{\chi} } \right)^{2}\chi \overline{c}c \equiv f_{\chi}^{c}\chi \overline{c}c .
	\end{equation}
Provided it is kinematically allowed, the decay rate into charmed mesons is approximately
	\bea
	\Gamma(\chi \to \overline{c}{c}) &\approx& \frac{3}{8\pi}(f_{\chi}^{c})^{2}m_{\chi}\left(1-\frac{4m_{D}^{2}}{m_{\chi}^{2}}\right)^{3/2} \nn \\
&\approx& 10^{-12} \; \mathrm{GeV}  \left( \frac{m_{\chi}}{10 \; \mathrm{GeV}} \right)\left( \frac{v_{\chi a} }{ 1 \; \mathrm{GeV} } \right)^{2} \left( \frac{ 1 \; \mathrm{TeV} }{ \Lambda_{\chi} } \right)^{4} \, .
	\eea
We impose the conservative requirement that it should not be abundantly produced to avoid any dilution of the baryon asymmetry, so we impose an early decay:
	\begin{equation}
	\label{eq:chidecayrate}
	\Gamma_{\chi} \gtrsim H(T = m_{\chi}) =\frac{1.66\sqrt{g_{\ast}}m_{\chi}^{2}}{M_{Pl}} \approx \bigg( \frac{m_{\chi}}{10 \; \mathrm{GeV}} \bigg)^{2} \times 10^{-17} \; \mathrm{GeV},
	\end{equation}
which is easy to achieve given the decay rate into $\overline{c}c$. From eq.~(\ref{eq:chidecayrate}) we find
	\begin{equation}
c\tau_{\chi}   \lesssim 1 \; \mathrm{m} \times  \bigg( \frac{10 \; \mathrm{GeV}}{m_{\chi}} \bigg)^{2}, 
	\end{equation}
meaning $\chi$ can decay at displaced vertices.
	
\subsubsection{Top quark decays}

The interactions of $\chi$ also induce exotic decays for the top quark, $t \to c \chi $ and $t \to c \chi \chi $. The corresponding partial widths are 
	\begin{eqnarray}
	\Gamma(t \to c \chi  )& \approx & \frac{ |\tilde{f}_{32}|^{2}  v_{\chi a}^{2}v_{\phi}^{2} m_{t}}{ 64\pi \Lambda_{\chi}^{4} } = |\tilde{f}_{32}|^{2} \bigg( \frac{v_{\chi a}}{1 \; \mathrm{GeV} } \bigg)^{2} \left( \frac{1 \; \mathrm{TeV} }{ \Lambda_{\chi}} \right)^{4} \times 1.3\times10^{-8} \; \mathrm{GeV} \, ,\\
	\Gamma(t \to  c \chi \chi) & \approx & \frac{ |\tilde{f}_{32}|^{2}v_{\phi}^{2}m_{t}^{3}}{12288\pi^{2}\Lambda_{\chi}^{4}} = |\tilde{f}_{32}|^{2}\left(\frac{1 \; \mathrm{TeV}}{\Lambda_{\chi}}\right)^{4}\times 2.5 \times 10^{-6} \; \mathrm{GeV} \, .
	\end{eqnarray}
We have ignored the final state masses in both expressions and suppressed the $\mathcal{O}(1)$ entries of the rotation matrices. Comparing with the top quark width $\Gamma_{t} = 1.4$ GeV~\cite{Agashe:2014kda,Abazov:2012vd,Khachatryan:2014nda}, we see the exotic branching fractions are highly suppressed.
We now discuss how these phenomenological predictions are altered when considering $Q_{FN}(\chi)=-1$.

\subsection{Models B-2: $Q_{FN}(\chi)=-1$ }
\begin{figure}[t]
\begin{center}
\includegraphics[width=250pt]{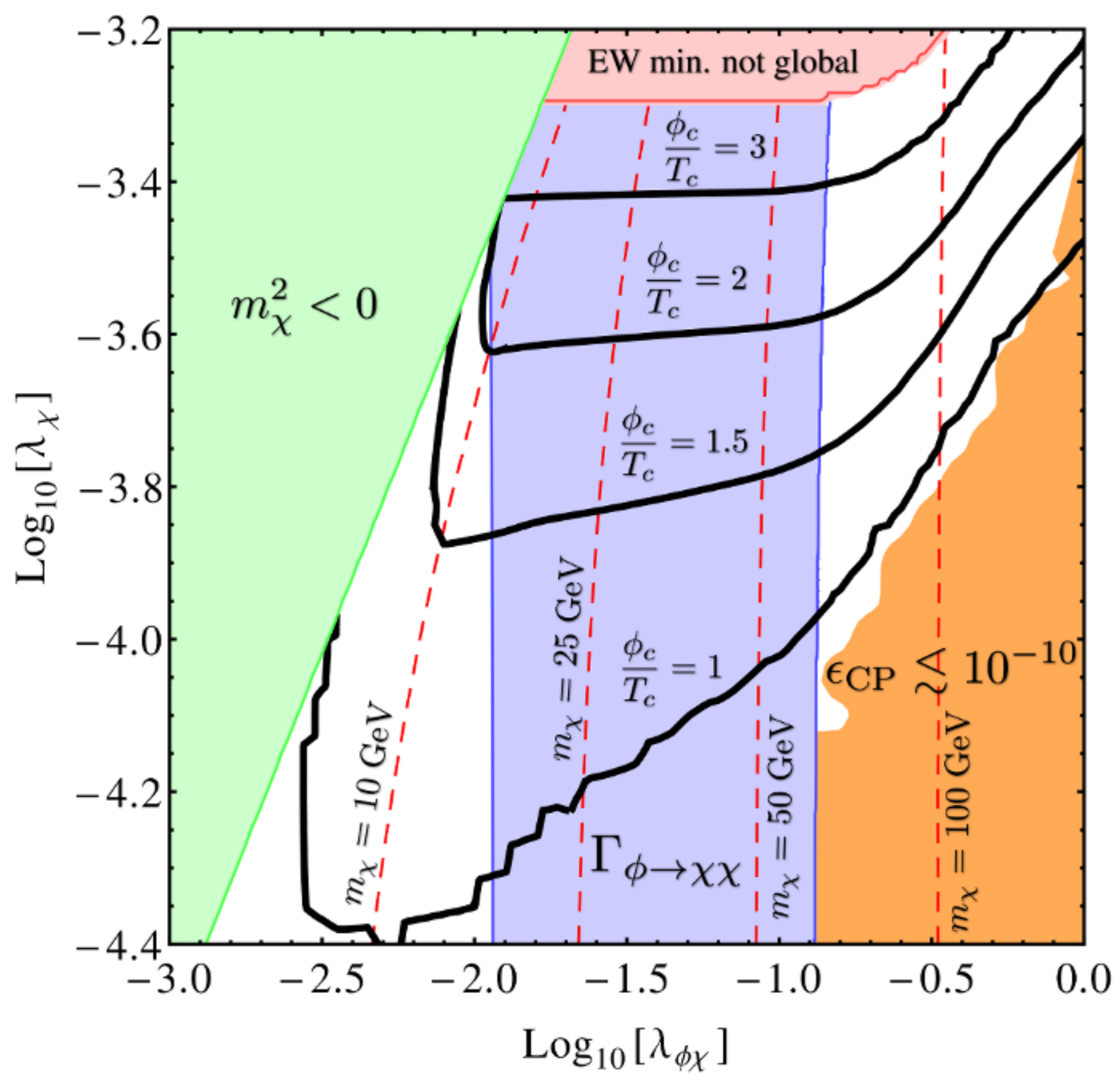}
\end{center}
\caption{\small Same parameter scan as in figure~\ref{fig:scan} but for model B-2 with $Q_{\rm FN}(X)=-1$.  We obtain strong first-order phase transitions for smaller values of $m_{\chi}$ than in model B-1, due to a larger effect from the varying Yukawa couplings.}
\label{fig:scan2}
\end{figure}

\begin{figure}[t]
\begin{center}
\includegraphics[width=250pt]{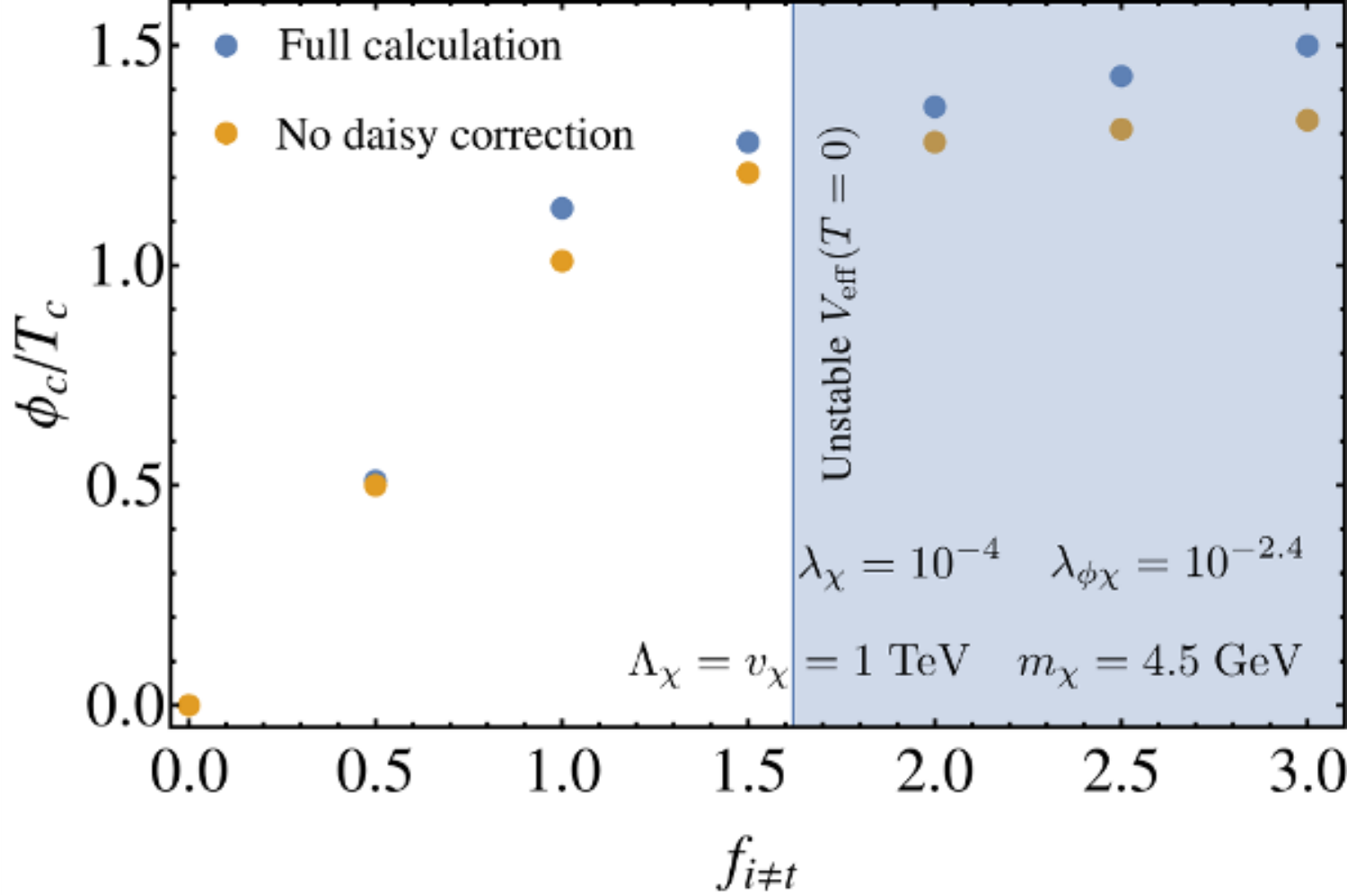}
\end{center}
\caption{\small Strength of the phase transition as a function of the Yukawa couplings (as in figure~\ref{fig:modb_yuk} but now for $Q_{\rm FN}(X)=-1$). The effective Yukawa couplings are now larger for given values of $f_{i}$ and $\chi$ than for $Q_{\rm FN}(X)=-1/2$ which results in stronger phase transitions when compared to figure~\ref{fig:modb_yuk}. However, for $f_{i\neq t}\gtrsim 1.6$, the $T=0$ effective potential is unstable in the $\phi$ direction along $\chi=\Lambda_{s}$, with no barrier present, which rules out such large couplings.}
\label{fig:modc_yuk}
\end{figure}

The scalar potential up to dimension four operators is now given by
	\begin{align}
\label{eq:2FieldPotC}
	V & = \mu_{\phi}^{2}\Phi^{\dagger}\Phi+\lambda_{\phi}(\Phi^{\dagger}\Phi)^{2} + \mu_{s}^{2}S^{\dagger}S+\lambda_{s}(S^{\dagger}S)^{2} + \mu_{\chi}^{2}X^{\dagger}X+\lambda_{\chi}(X^{\dagger}X)^{2}  \nonumber \\
	   & + \lambda_{\phi s}(\Phi^{\dagger}\Phi)(S^{\dagger}S) + \lambda_{\phi \chi}(\Phi^{\dagger}\Phi)(X^{\dagger}X) + \lambda_{\chi s}(X^{\dagger}X)(S^{\dagger}S) + \tilde \mu_{\chi s} (X^{\dagger}X^{\dagger}SS + H.c.) \\
	   & +\tilde{m_{\chi s}}(X^{\dagger}S+H.c)+\tilde{\lambda_{\chi s}}(S^{\dagger}S^{\dagger}SX+H.c.)+... \nonumber
	\end{align}
where the dots  indicate further $Z_{2}$ breaking terms. This symmetry is broken by the Froggatt--Nielsen Yukawa operators with odd exponents for the flavon. Therefore, in contrast with model B-1, even if we remove those terms by enforcing a $Z_2$ symmetry in the scalar potential, they would be regenerated at loop level. We will assume  these $Z_2$ breaking operators are kept small to avoid phenomenologically unacceptably large VEVs for $X$.
In this case,  we return to the same tree level potential as for model B-1. The phase transition proceeds in a similar way. However, the smaller powers in the exponents controlling the size of the effective Yukawas ensure these are larger than the effective Yukawas in model B for the same values of $\chi$. Hence the phase transition is strongly first-order for larger areas of parameter space. This also increases the CP violation. This is shown in figures~\ref{fig:scan2} and \ref{fig:modc_yuk}. Note that the choice $\Lambda_{\chi}=1$ TeV  is in tension with the constraints from Meson oscillations (see Appendix~\ref{sec:constraints2}). Here we keep $\Lambda_{\chi}=1$ TeV to allow easy comparison with Model B-1. We checked that strong phase transitions are also possible for $\Lambda_{\chi}=5$ TeV but at the cost of choosing much smaller quartic couplings.
 The constraints from the Higgs decay and stability of the potential (for $f_{i\neq t}=1$) remain the same, but overall, Model B-2 is under  higher experimental pressure than Model B-1.
One could study variants of these models by changing the FN charge assignments of the quarks, which can help alleviating the Meson oscillation constraints.

\subsubsection{Decays of $\chi$}
There is now no $Z_2$ symmetry in the Yukawa sector preventing $\chi$ from decaying. The leading contribution again comes from rotating the $\chi t' c'$ coupling into the mass basis
	\begin{equation}
	\mathcal{L} = \tilde{f}_{32} \left( \frac{X} {\Lambda_{\chi}} \right)\overline{t_{L}'}c_{R}' +H.c. \to \mathrm{Re}[U_{u}^{32 \ast}W_{u}^{22}\tilde{f}_{32}]\frac{ v_{\phi} }{ 2\Lambda_{\chi} } \chi \overline{c}c.
	\end{equation}
Following a similar analysis to Section~\ref{sec:chidecaychargehalf} above, we find a decay rate
	\begin{equation}
	\label{eq:chidecay}
	\Gamma(\chi \to \overline{c}{c}) \approx 10^{-5} \; \mathrm{GeV} \; \left( \frac{ m_{\chi} }{ 10 \; \mathrm{GeV} }\right) \left( \frac{ 1 \; \mathrm{TeV} }{ \Lambda_{\chi} }\right)^{2} .
	\end{equation}
This implies that $\chi$ decays promptly on collider and cosmological scales for typical parameter choices.

\subsubsection{Exotic top decay $t \to c \chi$}
The coupling  leading to exotic top decays $t \to c \chi$ is now less suppressed: $ v_{\phi}/(2\Lambda_{\chi}) \chi\overline{c}_Lt_R$ due to the larger flavor charge of $X$ compared to model B-1. The decay rate is given by
	\begin{equation}
	\Gamma(t \to c \chi) \approx \frac{|\tilde{f}_{32}|^{2}}{128 \pi }\left( \frac{ v_{\phi} }{ \Lambda_{\chi} } \right)^{2} m_{t} =  |\tilde{f}_{32}|^{2} \left( \frac{1 \; \mathrm{TeV} }{ \Lambda_{\chi} } \right)^{2} \times 2.6 \times 10^{-2} \; \mathrm{GeV}.
	\end{equation}
Comparing again to the measured top width $\Gamma_{t} = 1.4$ GeV~\cite{Agashe:2014kda,Abazov:2012vd,Khachatryan:2014nda}, we see that $\sim \mathcal{O}(1) \%$ branching fractions into the exotic state are now possible and therefore in the reach of LHC experiments~\cite{Aad:2015pja}. However, the final state would not necessarily be easy to distinguish from background as, from eq.~(\ref{eq:chidecay}), $\chi$ decays promptly to jets ($\overline{c}c$). We leave a more detailed study for further work.\

\section{CP violation and the baryonic yield}
\label{sec:CP}

The main motivation for this work is ultimately a natural setup for EW baryogenesis.
We have identified models in which Yukawa couplings can vary by values of order 1 during the EW phase transition. This not only can impact the nature of the EWPT but also generates large CP violating sources.
Calculation of the  baryon asymmetry produced during a first-order EW phase transition is 
intricate. It requires solving diffusion equations in front of the bubble wall after identification of the CP-violating source terms \cite{Cohen:1994ss}. This has been done in the supersymmetric context where the CP-violating source comes from the chargino mass matrix. However it has never been analysed in the context of a varying CKM matrix. We carry out a detailed comprehensive derivation in \cite{dynamicalCKMCP} and refer to this paper for more details.
For this work, it is enough to present a simplified estimate of the baryon asymmetry
based on the corresponding calculation in the one-flavor case. 

The CP violation in EW baryogenesis arises from the change of
the mass matrix along the
bubble wall in the first-order EWPT. In
principle, this is a purely kinematic effect
since particles with different mass profiles will experience the wall as
a potential barrier that they have
to surpass. In the case of fermions, this effect can even depend on the
helicity of the particle and violate CP. This requires that the Higgs
coupling to the particle (that produces its mass term) violates CP in
the first place.

In the one-flavor case, a mass term of the form
\be
m = |m| \, \exp(i \theta) \, ,
\ee
leads to the following source of CP violation in the
semi-classical Boltzmann equation
\be
\label{eq:CPV_one_flavor}
 \frac{1}{4} (\theta^\prime |m|^2)^\prime  \, .
\ee
This is the well-known CP-violating source found in the one-flavor model
in the semi-classical force approach~\cite{Cline:1997vk,Cline:2000nw, Kainulainen:2001cn}. In the case of several
flavors, the matter is much more complicated. In particular, the case of
chargino or neutralino driven electroweak baryogenesis in the MSSM was
the topic of long debate in the literature (see~\cite{Carena:2002ss, Konstandin:2005cd, Cirigliano:2006dg,Cirigliano:2009yd} and references therein). To simplify the
discussion,
we will neglect off-diagonal terms in the particle densities in the following.
This is for example a good approximation when there is a sizable mass splitting
between the eigenvalues of the system. Then, fast flavor oscillations
will erase the impact of the off-diagonal elements in the Green's
function (in the basis where the masses are diagonal).
It is also a good approximation in the limit where fast scatterings erase off-diagonal
densities quickly.
Still, mixing is important in the
forces, and one of the dominant mixing sources is of the form~\cite{Prokopec:2003pj,Prokopec:2004ic,Konstandin:2004gy,dynamicalCKMCP}
\be
\label{eq:CPV_multi_flavor}
{\rm Im} \left [W_{q}^\dagger m^{\dagger \prime \prime} m W_{q} \right]_{ii} \, ,
\ee
where $m$ denotes the mass matrix of fermion in question, $W_{q}$ is the
unitary matrix that is used to diagonalize $m^{\dagger} m$ and a prime
denotes the derivative with respect to the spatial coordinate across the
wall. This expression generalizes the one-flavor source in eq.~(\ref{eq:CPV_one_flavor}).

In the present model of Froggatt--Nielsen type, the CP violation arises
from the interplay of several flavors.
For example, the mass matrix of the top-charm sector resembles
\be
\label{eq:FNmass}
m  = v_{\phi} \left(
\begin{array}{cc}
c_{tt} \epsilon^0 & c_{tc} \epsilon^1 \\
c_{ct} \epsilon^2 & c_{cc}\epsilon^3 \\
\end{array}
\right) \, .
\ee
The VEV $\epsilon$ constitutes the source of breaking of the FN symmetry.
For the time being we assume it to be real-valued. The coefficients
$c_{ab}$ are complex constants. In particular, the complex coefficients
can give rise to CP-violating sources in the transport equation
according to (\ref{eq:CPV_multi_flavor}). In \emph{models A} all entries are of order
one in the symmetric phase, a sizable CPV source results in the
Boltzmann equations. Likewise, In \emph{models B} the Yukawa couplings involving the
second FN field $\chi$ can lead to a large mixing source.

It is essential to have Yukawa couplings of order unity, $\epsilon \simeq 1$, for sufficient
CP violation from the mass (\ref{eq:FNmass}). This can
be understood using an argument similar to the one given by Jarlskog in
the Standard Model. Due to the flavor structure, one can remove many of
the complex phases of the mass matrix (\ref{eq:FNmass}) using a constant
flavor basis change. In total, three of the four phases can be removed
and the remaining phase can be moved into one of the four constants
$c_{ab}$ at will. This shows that if one of the constants vanishes, CP
violation is absent. In turn, the CP violation is suppressed by a rather
large power of $\epsilon$.
Note that this argument is not quite the same as for the Standard
Model. In the present model, due to the changing $\epsilon$ many
combinations that are invariant under flavor basis changes can be
constructed that trivially vanish in the Standard Model due to $m^\prime
\propto m$. Hence, CP violation can be present in a two-flavor system when 
Yukawa couplings are varying.

Besides, if the FN symmetry breaking VEV $\epsilon$ has a complex
phase that is changing during the phase transition, one single quark flavor  can
provide a very strong source of CP violation according to (\ref{eq:CPV_one_flavor}). 
Again, one requires Yukawa couplings
of order unity in this case since otherwise the factor $\epsilon^3$ in the diagonal of (\ref{eq:FNmass}) leads to too much suppression.
In other words, without $\epsilon \sim 1$ in the symmetric phase, the bottom quark would be too light to lead to sufficient CP violation even with a changing phase across the wall.

\vskip 0.2 cm

In Table \ref{tab:bau}, we give some numerical results for the baryon
asymmetry $\eta / \eta_{obs}$ for the two sources
(\ref{eq:CPV_multi_flavor}) and (\ref{eq:CPV_one_flavor}).
The results for the mixing source are obtained by solving a diffusion
equation for two families of quarks.
The full analysis of the baryon asymmetry will be given in \cite{dynamicalCKMCP}. It is  striking that the correct amount of baryon asymmetry is reproduced. 

\begin{table}
\begin{center}
\begin{tabular}{| l | c | c | c || c |}
\hline
  Type & CPV source & $\phi_c/T_c$ & $l_w \, T_c$ & $\eta/\eta_{obs}$ \\
\hline
\hline
One-flavor case (\ref{eq:CPV_one_flavor}) &
  $\Delta \theta = 0.2 $ &
  1.6 & 6 & $\simeq 1.0$ \\
\hline
Two-flavor  case (\ref{eq:CPV_multi_flavor}) and (\ref{eq:FNmass}) &
  $ \arg(c_{tt}) = \pi/2$ &
  1.0 & 8 & $\simeq 2.6$ \\
\hline
\end{tabular}
\end{center}
\caption{\small The baryon asymmetry for two benchmark points corresponding to two
different CP-violating sources. The relevant parameters are the Higgs VEV during the phase transition $\phi_c$, the wall thickness $l_w$ and the CP-violating complex phase. }
\label{tab:bau}
\end{table}

\section{Discussion and Conclusion}

The cosmology of flavor physics had so far not be studied.
An important implication is the possibility of exploiting dynamical Yukawa couplings 
for EW baryogenesis\footnote{An obvious other direction of research would be to apply the idea in the leptonic sector and consider implications for leptogenesis.}. This can be done in the several few scenarios addressing the flavor problem, namely Froggatt--Nielsen models, Randall-Sundrum models and Composite Higgs models.
This paper is a first investigation in this direction, focusing on FN models as a benchmark scenario.
The aim of this study was to identify classes of FN models which can provide the  dynamics leading to variable quark Yukawa couplings during the EW phase transition while satisfying all experimental bounds. 
The summary of our findings is the following:

In the first class of models, the flavon $\sigma$ has to be ultra light, $m_{\sigma} \lesssim 1$ GeV, to induce  Yukawa   
coupling variation during the EW phase transition which, as a result of this variation,  is first-order.
The Yukawa coupling variation also induces  large enough CP violation for generating the baryon asymmetry of the universe.
This model can be compatible with flavor constraints  if the alignment mechanism advocated in \cite{Knapen:2015hia} is at work.  However, there are tuning issues related to the radiative corrections of the flavon mass.

In the second class of models, the tuning issues are alleviated. There are two scalar flavons. One is heavy, $\sigma$, at the TeV scale, together with the heavy FN vector-like fermions, and the other one, $\chi$,  is light, in the 10--70 GeV range.
This light flavon has a sizable mixing with the Higgs. The Higgs can decay 
$\phi \to \chi \chi$ where the leading decay of $\chi$ is $\chi \to c \bar{c}$. Another specific signature of this model is the exotic top quark decay $t \to c  \chi$.
Because the FN scale is low, $\Lambda_s \sim 1$ TeV, new fermions at the TeV scale are expected.
LHC searches for vector-like FN quarks are thus an important test of this model.
In this second class of models, the region of parameter space where the Yukawa variation is responsible for the first-order phase transition is rather limited. In most of the parameter space, the first-order phase transition comes from a tree level barrier in the $(\phi, \chi)$ scalar potential. The Yukawa variation during the EWPT coming from the $\chi$ field  is on the other hand responsible for large CP-violating sources guaranteeing a large enough baryon asymmetry.

In conclusion, 
the main generic prediction 
is the existence of a light (below EW scale) additional scalar with observable signatures at the LHC.
This follows from the requirement that the flavon VEV should vary by a value of order the FN scale (at least TeV as imposed by flavor constraints)
when the Higgs acquires its VEV, for a successful impact on EW baryogenesis.
 This demands a rather flat potential in the flavon direction, in order for the flavon VEV to change by a value of order $ \Lambda_{s} \sim$ TeV, when the Higgs acquires a VEV of order $ v_{\phi}$.
This light flavon is the main obstacle in our constructions as it typically clashes with flavour constraints.
The requirement of the light flavon comes from our equation (\ref{eq:minimum}) 
which follows from our simple assumption for the form  of the Higgs-Flavon scalar potential.

We do not think that the presence of this light flavon 
  is a generic prediction of models of varying Yukawas at the EW scale.
   For instance, in the context of Randall--Sundrum models where the origin of the fermion mass hierarchy is of a very different nature, varying Yukawas correlated with the EW phase transition turn out to be easier to implement (i.e. less constrained by experiments) and in fact quite natural~\cite{RSdynamicalCKMCP}.
 In this case,  the flavon (played by the radion/dilaton in \cite{RSdynamicalCKMCP}) has its own dynamics which induces EW symmetry breaking in such a way that the Higgs mass parameter is proportional to the flavon VEV. This enables the possibility to have Yukawa coupling variation during the EW phase transition compatible with flavour constraints as the flavon is parametrically heavier than the Higgs, in the TeV range.
It will be interesting to investigate other realizations of varying Yukawas during the EW phase transition along this line, as they are flavor motivated incarnations of the EW baryogenesis mechanism.

\subsubsection*{Acknowledgements}
We acknowledge support by the German Science Foundation (DFG) within the Collaborative Research Center (SFB) 676 Particles, Strings and the Early Universe.

\appendix

\section{Flavon couplings in the mass basis}
\label{sec:flavoncouplings}
In this appendix we clarify the transformations between the flavor and mass basis.
\subsection{Interactions with one flavon field}

We begin with the flavon couplings
		\begin{align}
	\mathcal{L} & \supset \tilde{y_{ij}}\left(\frac{S }{ \Lambda_{s} }\right)^{\tilde{n}_{ij}} \overline{Q_{i}'}  \tilde{\Phi} U_{j}'  + y_{ij}\left(\frac{S }{ \Lambda_{s} }\right)^{n_{ij}} \overline{Q'}_i \Phi D_{j}'   + H.c.,
	\end{align}  
where we now explicitly denote quark fields in the flavor basis with primes. After symmetry breaking the couplings become
	\begin{align}
	\mathcal{L} & \supset \frac{v_{\phi}}{\sqrt{2}}\left(1+\frac{\phi}{v_{\phi}} + \tilde{n}_{ij}\frac{\sigma}{v_{s}}\right)\tilde{y_{ij}}\epsilon_{s}^{\tilde{n}_{ij}}  \overline{u_{Li}'} u_{Rj}' \\
		    &  + \frac{v_{\phi}}{\sqrt{2}}\left(1+\frac{\phi}{v_{\phi}} + n_{ij}\frac{\sigma}{v_{s}}\right)y_{ij}\epsilon_{s}^{n_{ij}} \overline{d_{Ri}'} d_{Lj}'   + H.c.
	\end{align}  
Defining $Y_{ij} \equiv  y_{ij}\epsilon_{s}^{n_{ij}}$ and $\tilde{Y_{ij}} \equiv  \tilde{y_{ij}}\epsilon_{s}^{\tilde{n}_{ij}}$, we can diagonalize the Yukawa coupling matrices as in the SM~\cite{Peskin:1995ev}
	\begin{align}
	& (Y_{u})^{2}=U_{u}^{\dagger}\tilde{Y}\tilde{Y}^{\dagger}U_{u}, \quad (Y_{u})^{2}=W_{u}^{\dagger}\tilde{Y}^{\dagger}\tilde{Y}W_{u}, \\
	& (Y_{d})^{2}=U_{d}^{\dagger}YY^{\dagger}U_{d}, \quad (Y_{d})^{2}=W_{d}^{\dagger}Y^{\dagger}YW_{d},
	\end{align}
where $U_{u/d}$ and $W_{u/d}$ are unitary matrices and $Y_{u}$ and $Y_{d}$ are diagonal. Consequently one also has
	\begin{equation}
	Y_{u} = U_{u}^{\dagger}\tilde{Y}W_{u}, \quad Y_{d} = U_{d}^{\dagger}YW_{d}.
	\end{equation}
The CKM matrix is given by
	\begin{equation}
	V_{\rm CKM} = U_{u}^{\dagger}U_{d}.
	\end{equation}
Applying the transformations
	\begin{equation}
	u_{L}'=U_{u}u_{L}, \quad  u_{R}'= W_{u}u_{R}, \quad d_{L}'=U_{d}d_{L}, \quad  d_{R}'= W_{d}d_{R},
	\end{equation}
we diagonalize the mass terms and Higgs couplings and obtain
	\begin{equation}
	\mathcal{L} \supset m_{u}^{i}\overline{u_{Li}}u_{Ri}\left(1+\frac{\phi}{\sqrt{2}}\right)+ m_{d}^{i}\overline{d_{Li}}d_{Ri}\left(1+\frac{\phi}{\sqrt{2}}\right)+H.c. \, ,
	\end{equation}
where $m_{u}^{i} = Y_{u}^{i}v_{\phi}/\sqrt{2}$ and $m_{d}^{ii} = Y_{d}^{i}v_{\phi}/\sqrt{2}$. In the mass basis the flavon interactions become~\cite{Huitu:2016pwk}
	\begin{equation}
	\mathcal{L} \supset \frac{1}{\sqrt{2}}\frac{v_{\phi}}{v_{s}}\sigma \left(\overline{u_{L}}U_{u}^{\dagger}(\tilde{n}\tilde{Y})W_{u}u_{R}+\overline{d_{L}}U_{d}^{\dagger}(nY)W_{d}d_{R} \right)+H.c. \, ,
	\end{equation}
where $(\tilde{n}\tilde{Y})_{ij} = \tilde{n}_{ij}\tilde{Y}_{ij}$ and  $(nY)_{ij} = n_{ij}Y_{ij}$. Note the matrices which rotate the Yukawa couplings to bring us into the mass basis also rotate the flavon couplings. However, because of the $n_{ij}$ factor, these rotations do not bring the flavon couplings into a diagonal form. The flavon couplings, $C_{u} \equiv (v_{\phi}/v_{s})U_{u}^{\dagger}(\tilde{n}\tilde{Y})W_{u}$ and $C_{d}\equiv (v_{\phi}/v_{s})U_{d}^{\dagger}(nY)W_{d}$ can be approximated by
	\begin{align}
	C_{u}^{ij} = (1+\delta_{ij}[\tilde{n}_{ij}-1])\epsilon_{s}^{\tilde{n}_{ij}}\frac{v_{\phi}}{v_{s}}, \\
	 C_{d}^{ij} = (1+\delta_{ij}[n_{ij}-1])\epsilon_{s}^{n_{ij}}\frac{v_{\phi}}{v_{s}}.
	\end{align}
In this approximate form the factors $n_{ij}$ do not appear in the off-diagonal couplings~\cite{Bauer:2016rxs}. The rotation helps to reduce the magnitude of these entries. To see intuitively why this occurs, note that, if all $n_{ij}$ were the same, the rotation would bring the flavon couplings into diagonal form, erasing these entries.

\subsection{Higher order interactions}
\label{sec:higheroderflavon}

The original Lagrangian also contains higher order flavon interactions. These are particularly important for model B, in which the $\chi$ flavon has a suppressed VEV. After symmetry breaking, the interactions with no $v_{\chi}$ insertions become
	\begin{equation}
	\mathcal{L} \supset \frac{v_{\phi}}{\sqrt{2}} \left(\overline{u_{L}}U_{u}^{\dagger}(\tilde{F}\chi^{\tilde{n}})W_{u}u_{R}+\overline{d_{L}}U_{d}^{\dagger}(F\chi^{n})W_{d}d_{R} \right)+H.c. \, ,
	\end{equation}
where the rotation now acts on the matrices with entries
	\begin{align}
	(\tilde{F}\chi^{\tilde{n}})_{ij} = \tilde{f_{ij}}\left(\frac{\chi}{\sqrt{2}\Lambda_{\chi}}\right)^{\tilde{m}_{ij}},	\\
	(F\chi^{n})_{ij} = f_{ij}\left(\frac{\chi}{\sqrt{2}\Lambda_{\chi}}\right)^{m_{ij}}.
	\end{align}
Hence the dimension of the interaction is also mixed when going from the flavor to the quark mass basis.

 \section{Experimental Constraints on a light flavon}
 \label{section:flavorconstraints}
 
 The nature of the constraints on the flavon $\sigma$ depends on whether it  is heavier or lighter than the neutral kaon $K$. We denote $m_{\sigma} > M_{K}$ ($m_{\sigma} < M_{K}$) the heavy (light) scalar case.
 
\subsection{Heavy scalar case, $m_{\sigma} > M_{K}$}
\begin{figure}[t]
\begin{center}
\includegraphics[width=225pt]{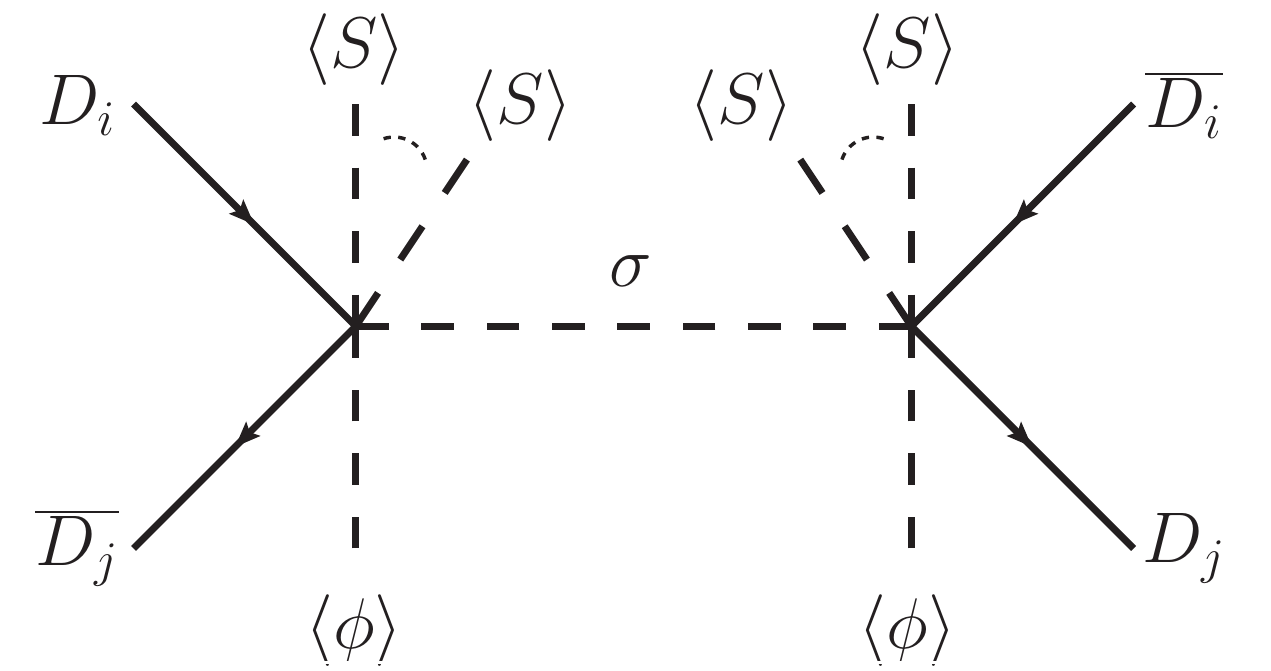}
\end{center}
\caption{\small Diagram contributing to neutral meson oscillations. The number of $\langle S \rangle$ insertions depends on the FN charges of the quarks. Each VEV insertion leads to a suppression of order $\epsilon_{s}=\langle S \rangle / \Lambda_{s}$.}
\label{fig:mesons_s}
\end{figure}

The off diagonal terms allowed by the FN charges in the Yukawa couplings generate the CKM mixing. They also lead to  tree level flavor changing neutral currents (FCNCs) such as neutral meson oscillations as shown in figure~\ref{fig:mesons_s}. For example, consider the terms relevant for $K-\overline{K}$ oscillations
	\begin{align}
	\mathcal{L} \; = & \; \left(\frac{S}{\Lambda_{s}}\right)^{5}\overline{Q_{2L}} \Phi D_{1R} + \left(\frac{S}{\Lambda_{s}}\right)^{5}\overline{Q_{1L}} \Phi D_{2R}  + H.c. \nonumber \\
		   \; \to & \; \epsilon_s^{4} \frac{ v_{\phi} }{ 2\Lambda_{s} } \sigma \Big(  \overline{s}Rd +  \overline{d}Ls +  \overline{d} R s +  \overline{s} R d \Big), 
	\end{align}
where the second line is generated after spontaneous symmetry breaking, L and R are the usual left and right projection operators and we have suppressed the $\mathcal{O}(1)$ coefficients. Such terms induce Wilson coefficients at tree level which contribute to the effective Hamiltonian
	\begin{equation}
	\mathcal{H} = C_{2}^{sd}(\overline{s}Ld)^{2}+\tilde C_2^{sd}(\overline{s}Rd)^{2}+C_{4}^{sd}(\overline{s}Ld)(\overline{s}Rd)+H.c.
	\end{equation}
The Wilson coefficients are given by
	\begin{align}
	C_{2}^{sd} = \left( \frac{\epsilon_{s}^4 v_{\phi}}{ 2\Lambda_{s} m_{\sigma}} \right)^{2}, \qquad
	\tilde C_2^{sd} =\left( \frac{\epsilon_{s}^4 v_{\phi}}{ 2\Lambda_{s} m_{\sigma}} \right)^{2}, \qquad
	C_{4}^{sd} = \left( \frac{\epsilon_{s}^4 v_{\phi}}{ 2\Lambda_{s} m_{\sigma}} \right)^{2} \, .
	\end{align}
Limits on these Wilson coefficients have been derived by the UTfit collaboration~\cite{Bona:2007vi} (we use the updated limits presented in~\cite{Derkach}). The most stringent constraint is set on Im$[C_{4}^{sd}]$ and reads $|\mathrm{Im}[C_{4}^{sd}]| \lesssim 3.8\times10^{-18} \; \text{GeV}^{-2}$. This gives a limit
	\begin{equation}
	\label{eq:epskaon}
	\sqrt{\Lambda_{s}m_{\sigma}} \gtrsim 10 \; \text{TeV} \qquad ( \; \text{from Im}[C_{4}^{sd}] \;).
	\end{equation}
The other  constraints are
\bea
	\label{eq:rc4sd}
	\sqrt{\Lambda_{s}m_{\sigma}} &\gtrsim & 2.4 \; \text{TeV} \qquad ( \; \text{from Re}[C_{4}^{sd}] \;), \\
	\sqrt{\Lambda_{s}m_{\sigma}} &\gtrsim  &6.7 \; \text{TeV} \qquad ( \; \text{from Im}[C_{2}^{sd}] \;), \\
	\label{eq:ic2sd}
	\sqrt{\Lambda_{s}m_{\sigma}}& \gtrsim &1.7 \; \text{TeV} \qquad ( \; \text{from Re}[C_{2}^{sd}] \;).
	\label{eq:rc2sd}
	\eea
The limits on $C_{2}^{sd}$ also apply to $\tilde C_2^{sd}$, so similar limits apply to other choices of Yukawas, but we do not quote them here. We have also ignored renormalization group evolution which will mix the various Wilson coefficients below the scale of $m_{\sigma}$. From the discussion above we can see however, that there is a broadly applicable constraint, $\sqrt{\Lambda_{s}m_{\sigma}} \gtrsim$ few TeV, for $\mathcal{O}(1)$ Yukawas, even if some of the dimensionless coefficients are smaller than unity, which may lower some of the individual bounds. Limits of $\sqrt{\Lambda_{s}m_{\sigma}} \gtrsim$ few TeV are also obtained from Wilson coefficients contributing to $B-\overline{B}$ and $D-\overline{D}$ mixing~\cite{Bona:2007vi,Derkach}. 

\subsection{Light scalar case,  $m_{\sigma} < M_{K}$}
We now consider the possibility  $m_{\sigma} < M_{K}$. Again, $\sigma$ will mediate tree level neutral meson oscillations. A simple estimate of the resulting constraint is obtained by replacing the propagator
	\begin{equation}
	\frac{1}{q^{2}-m_{\sigma}^{2}} \to \frac{1}{m_{K}^{2}},
	\end{equation}
as $q^2\gg m_{\sigma}^2$. The constraint eq.~(\ref{eq:epskaon}) now becomes
	\begin{equation}
	\Lambda_{s} \gtrsim  10^{8} \; \text{GeV} \qquad ( \; \text{Im}[C_{4}^{sd}] \;).
	\end{equation}
Again, the other mixing effects give similar constraints. So we take 
\be
\Lambda_{s} \gtrsim 10^{8} \mbox{ GeV} 
\label{eq:mesonconstraint}
\ee
 as a bound from neutral meson oscillations in the light scalar case.

\begin{figure}[t]
\begin{center}
\includegraphics[width=400pt]{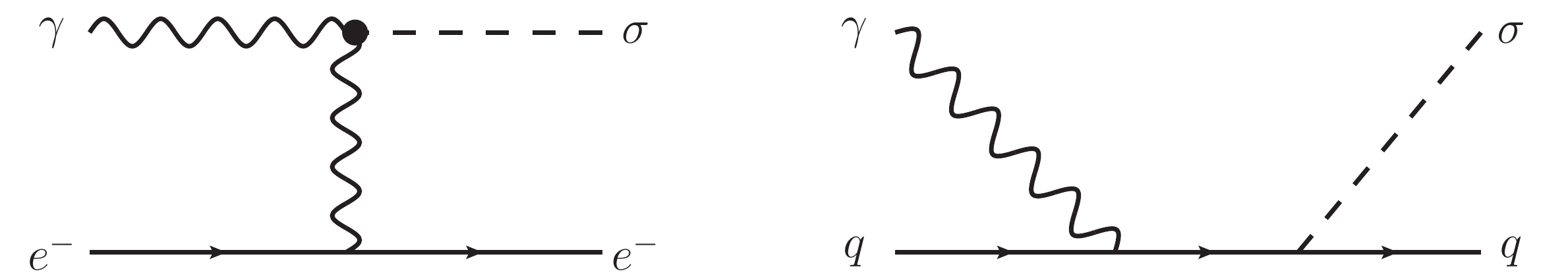}
\end{center}
\caption{\small Left: Primakoff process. Right: inelastic scattering mediated through the effective Yukawa. Similar diagrams exist with the electron replaced by a quark and photons replaced with gluons for both processes.}
\label{fig:eqscatterings}
\end{figure}

We now turn our attention to cosmological constraints. We saw in Section ~\ref{sec:treelevelmodelA}, eq.~(\ref{eq:sigmamasslim}), that for $\sigma$ to change from $\Lambda_{s}$ to $0.2\Lambda_{s}$ during the EWPT, this requires 
	\begin{equation}
	\label{eq:lightscalarsimplemass}
	m_{\sigma} \lesssim \frac{\sqrt{\lambda_{\phi}}v_{\phi}^{2}}{\Lambda_{s}}, \; \text{}
	\end{equation}
i.e. $m_{\sigma}\lesssim 0.2$ MeV for $\Lambda_{s} \gtrsim 10^{8}$ GeV. After the EWPT, loops of SM quarks generate a term in the Lagrangian of the form
	\begin{equation}
	\mathcal{L} \sim \frac{\sigma}{\Lambda_s} F_{\mu \nu}F^{\mu \nu} \, .
	\end{equation}
 This induces a decay of the flavon $s \to \gamma \gamma$, which is the only decay channel open for such a light scalar. The decay rate is given by
	\begin{equation}
	\Gamma(\sigma \to \gamma \gamma) \sim \frac{m_{\sigma}^{3}}{\Lambda_s^2}.
	\end{equation}
Limits on light scalars interacting with and decaying to photons have been derived in refs.~\cite{Masso:1995tw,Masso:1997ru,Cadamuro:2011fd}. The cosmological evolution of the $\sigma$ number density in our scenario proceeds as follows. Above the EWPT $\sigma$ comes into thermal abundance through scatterings such as $t\overline{c} \leftrightarrow \sigma \phi$. Once the top gains mass at the EWPT, this process rapidly drops out-of-equilibrium. The dominant interactions are then the Primakoff process (both gluon and photon initiated) and inelastic scattering mediated through the effective Yukawa of $\sigma$, $y_{\rm eff} \sim m_{q}/\Lambda_s$,  shown in figure~\ref{fig:eqscatterings}. Above $T\sim \Lambda_{\rm QCD}$ the gluon initiated Primakoff process has a rate
	\begin{equation}
	\Gamma_{A} \sim \frac{\alpha_{s}T^{3}}{\Lambda_{s}^{2}} \quad \text{for}  \quad T \gtrsim \Lambda_{\rm QCD},
	\end{equation}
where $\alpha_{s}$ is the strong fine structure constant. Hence the Primakoff process drops out of equilibrium at a temperature
	\begin{equation}
	T_{\rm{FO}} \sim 0.1 \; \mathrm{GeV} \; \left(\frac{\Lambda_{s}}{10^{8} \; \mathrm{GeV} } \right)^{2}.
	\end{equation}
Below $T\sim \Lambda_{\rm QCD}$, the rate is rapidly suppressed by the falling initial state nucleon or pion abundance and the confinement of color charge. The photon initiated Primakoff process is slightly weaker above $\Lambda_{\rm QCD}$ but is not suppressed below it. 

Above $T\sim \Lambda_{\rm QCD}$ the gluon initiated inelastic scattering process has a rate
	\begin{equation}
	\Gamma_{B} \sim \alpha_{s} \left(\frac{m_{b}}{\Lambda_{s}}\right)^{2}T \quad \text{for}  \quad T \gtrsim \Lambda_{\rm QCD} \, ,
	\end{equation}
and comes \emph{into} equilibrium at a temperature
	\begin{equation}
	T_{\rm{EQ}} \sim 100 \; \mathrm{GeV} \; \left(\frac{10^{8} \; \mathrm{GeV} }{\Lambda_{s}} \right)^{2} \, .
	\end{equation}
Below $T\sim \Lambda_{\rm QCD}$, the rate is rapidly suppressed by the massive propagator, the falling initial state nucleon or pion abundance and the confinement of color charge. Again there is also a slightly weaker photon initiated process, which is now also suppressed below $T\sim \Lambda_{\rm QCD}$, because of the falling initial state abundance.

Combining these estimates with the mass relation~(\ref{eq:lightscalarsimplemass}) means that in our scenario $\sigma$ always freezes out at $T_{\rm FO} \gg m_{\sigma}$. Hence the limits of ref.~\cite{Cadamuro:2011fd} apply (the additional interactions in this scenario result in a larger freeze-out abundance, $Y_{\rm FO}$, which work to make the limits stronger, compared to the case considered in~\cite{Cadamuro:2011fd}, in which the only interaction considered is the photon initiated Primakoff effect). Light scalars decaying to photons can e.g., disturb BBN by changing $n_{B}/n_{\gamma}$, decrease the effective number of neutrinos by preferentially heating the electron/photon bath, distort the CMB, lead to excess extragalactic background light and alter the ionization history of primordial hydrogen. Applying the constraints derived by the detailed study in~\cite{Cadamuro:2011fd} shows that our light scalar is constrained to a region of parameter space where 
\be
\Lambda_{s}\gtrsim 10^{12} \mbox{ GeV and } m_{\sigma} \lesssim 100 \mbox{ eV} \, .
\ee
 For even lighter scalars, $m_{\sigma}\lesssim 1$ eV, there are also bounds from long range forces~\cite{Raffelt:2012sp}, but we do not consider this range of parameter space here. 

\section{Constraints on a second FN field with negligible VEV today}
\label{sec:constraints2}
\subsection{Constraints on $\Lambda_{\chi}$ for $Q_{\rm FN}(X)=-1$ (Model B-2)}

\begin{figure}[t]
\begin{center}
\includegraphics[width=225pt]{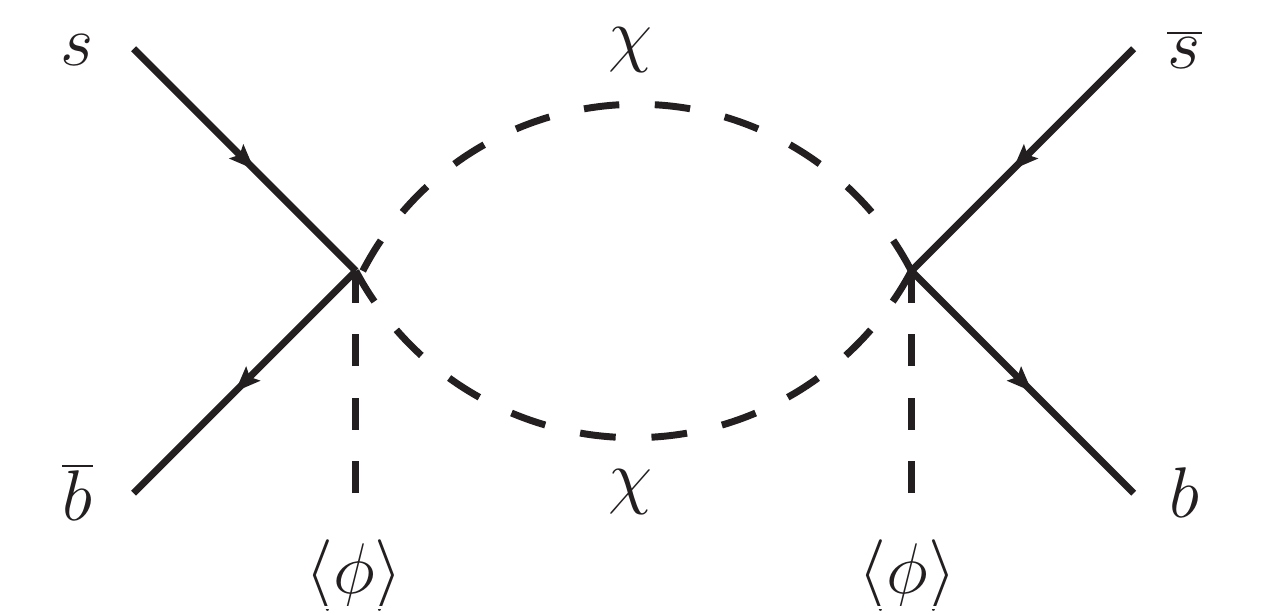}
\end{center}
\caption{\small One loop diagram leading to $B_{s}-\overline{B_{s}}$ oscillations for $Q_{\rm FN}(X)=-1$.}
\label{fig:mesons_x1}
\end{figure}

In model B, if the second scalar  $\chi$ has zero (or negligible) VEV $v_{\chi}$ today, the above tree level constraints do not apply. We therefore  examine now the loop level constraints. The  limits on the $\chi$ sector are weakened as VEV insertions are replaced by loop factors. 

We first examine the constraints for $Q_{\rm FN}(X)=-1$, assuming again a mass splitting arising from a source of explicit symmetry breaking, giving $m_{\chi} < m_{\eta}$. At one-loop level, $B_{s}-\overline{B_{s}}$ oscillations come from the exchange of $\chi$ and $\eta$, as depicted in figure~\ref{fig:mesons_x1}. The integral is effectively regulated by $\eta$ (if $m_{\eta}>\Lambda_{\chi}$, we should simply cut-off the integral at $\Lambda_{\chi}$). The contribution to the $C_{2}^{bs}$ Wilson coefficient is estimated to be
	\begin{equation}
	|C_{2}^{bs}| \approx \left(\frac{v_{\phi}}{2\Lambda_{\chi}^{2}}\right)^{2}\left\{ \frac{1}{(4\pi)^{2}}\left(\frac{1}{2}\frac{m_{\chi}^{2}+m_{\eta}^{2}}{m_{\chi}^{2}-m_{\eta}^{2}}\mathrm{Log}\left[\frac{m_{\chi}^{2}}{m_{\eta}^{2}} \right]  -1 \right) \right \},
	\end{equation} 
where the term in the curly brackets comes from  the loop integral. Note that, as required, this term approaches zero as $m_{\chi} \to m_{\eta}$. The contribution from the integral typically takes on values $\sim0.01$. The constraint on the Wilson coefficients reads $|C_{2}^{bs}|<3.8\times10^{-12}$ GeV$^{-2}$~\cite{Derkach}. This gives us a constraint
	\begin{equation}
	\label{eq:oneloopchi}
	\Lambda_{\chi} \gtrsim   2 \; \mathrm{TeV} \qquad ( \; \text{from} \; |C_{2}^{bs}| \;).
	\end{equation}
After rotation to the mass basis (see Appendix~\ref{sec:higheroderflavon}), one finds that a similar one-loop process also contributes to $K-\overline{K}$ oscillations. Taking into account the suppression of the coupling by $\sim \epsilon_{s}^{3}$ from the rotation, one finds a similar limit as in eq.~(\ref{eq:oneloopchi}). Another important constraint comes from $D-\overline{D}$ oscillations at tree level. The reason is that the rotation from the flavor into the mass basis leads to a coupling
	\begin{equation}
	\label{eq:twotermsfromrotation}
 	\tilde{f_{32}} \frac{v_{\phi}}{2\Lambda_{\chi}}\chi \overline{t_{L}'}c_{R}' \to \tilde{f_{32}}\frac{v_{\phi}}{2\Lambda_{\chi}}U_{u}^{32 \ast}W_{u}^{21} \chi \overline{c_{L}}d_{R}+\tilde{f_{32}}U_{u}^{31 \ast}W_{u}^{22}\frac{v_{\phi}}{2\Lambda_{\chi}} \chi \overline{d_{L}}c_{R}.
	\end{equation}
Numerically we find the typical suppression of the coupling from the rotation is $\sim \epsilon_{s}^{4}$. Comparing to the experimental constraint, $|C_{2}^{cu}| < 1.6\times10^{-13}$ GeV$^{-2}$~\cite{Derkach}, one finds a limit
	\begin{equation}
	\sqrt{\Lambda_{\chi}m_{\chi}} \gtrsim 500 \; \mathrm{GeV} \qquad ( \; \text{from} \; |C_{2}^{cu}| \;).
	\end{equation}
Note the qualitative different form of this limit means it particularly constraining on this scenario. The limit can be weakened if there is an accidental cancellation in the rotation, or if the pseudoscalar mass is close to $m_{\chi}$. We must assume one (or both) of these in order to have $\Lambda_{\chi}\sim 1$ TeV with $m_{\chi} \sim 10$ GeV in model B-2. (The presence of a light pseudoscalar is expected to have only negligible effect of the analysis of the phase transition.) Numerically a cancellation in the rotation at the $\sim 10$ \% level is sufficient. Alternatively this constraint can be evaded if $X$ couples only to the bottom type quarks, or if the pattern of charges is altered so that there are no single power of $X$ coupling to up type quarks~\cite{Bauer:2016rxs}. In both cases that strength of the phase transition is not expected to be altered much, but the $\chi$ and exotic top decays, discussed above, will be weakened (or possibly non-existent for the top).

At two-loop level we have a contribution to the $C_{2}^{bd}$ Wilson coefficient, which is constrained to be $|C_{2}^{bd}|<2.7\times10^{-13}$ GeV$^{-2}$~\cite{Derkach}. We approximate the two loop contribution to the Wilson coefficient as
	\begin{equation}
	|C_{2}^{bd}| \approx \left(\frac{v_{\phi}}{4\Lambda_{\chi}^{4}}\right)^{2}\left\{ \frac{1}{(16\pi)^{2}} \Lambda_{\chi}^{4} \right\} \, .
	\end{equation}
Taking the integral to be cut-off at $\Lambda_{\chi}$, we find a constraint
	\begin{equation}
	\Lambda_{\chi} \gtrsim  1.5 \; \mathrm{TeV} \qquad ( \; \text{from} \; |C_{2}^{bd}| \;).
	\end{equation}
The three-loop diagram contributing to $D-\overline{D}$ oscillations gives a weaker bound. However, there is a dangerous four-loop integral contributing to $K-\overline{K}$ oscillations.
This leads to a constraint
	\begin{equation}
	\Lambda_{\chi} \gtrsim 2.5 \; \mathrm{TeV} \qquad ( \; \text{from} \; |C_{4}^{sd}| \;).
	\end{equation}	
Given some of these constraints can be suppressed by the dimensionless couplings, we take a general constraint $\Lambda_{\chi} \gtrsim 1$ TeV for $Q_{\rm FN}(X)=-1$ in our analysis of this scenario.

\subsection{Constraints on $\Lambda_{\chi}$ for $Q_{\rm FN}(X)=-1/2$ (Model B-1)}

\begin{figure}[t]
\begin{center}
\includegraphics[width=225pt]{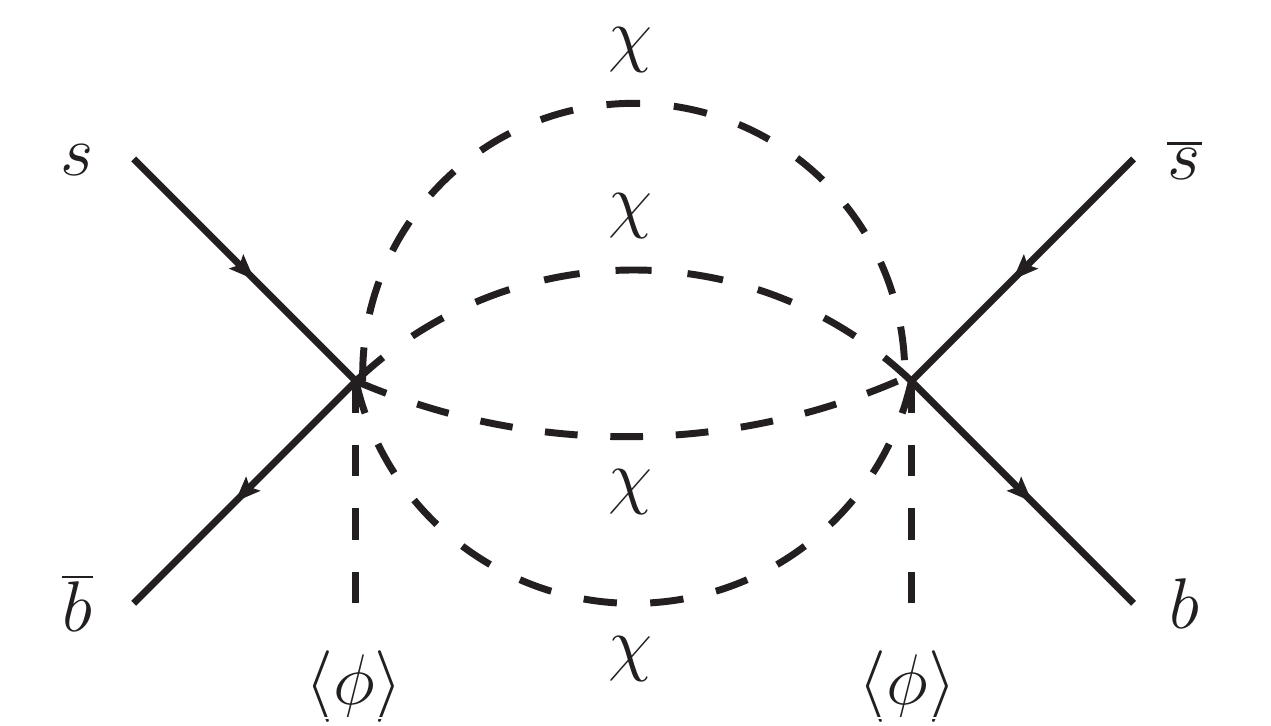}
\end{center}
\caption{\small Analagous diagram to figure~\ref{fig:mesons_x1} but for $Q_{\rm FN}(X)=-1/2$. The lowest level diagram occurs at three loop and leads to $B_{s}-\overline{B_{s}}$ oscillations.}
\label{fig:mesons_x}
\end{figure}

The constraints are even more relaxed for $Q_{\rm FN}(X)=-1/2$ as the Wilson coeffcients are generated at higher loop level.  $B_{s}-\overline{B_{s}}$ oscillations now occur at three-loop level, as depicted in figure~\ref{fig:mesons_x}. By cutting off the loop integrals at $\Lambda_{\chi}$, the contribution to the $C_{2}^{bs}$ Wilson coefficient is estimated as
	\begin{equation}
	|C_{2}^{bs}| \approx \left(\frac{v_{\phi}}{2^{5/2}\Lambda_{\chi}^{4}}\right)^{2}\left\{ \frac{1}{(16\pi)^{3}} \Lambda_{\chi}^{4} \right\},
	\end{equation} 
where the term in the curly brackets is our estimate of the tree-loop contribution. The constraint on the Wilson coefficient reads $|C_{2}^{bs}|<3.8\times10^{-12}$ GeV$^{-2}$~\cite{Derkach}, leading to the constraint
	\begin{equation}
	\Lambda_{\chi} \gtrsim  250 \; \mathrm{GeV} \qquad ( \; \text{from} \; |C_{2}^{bs}| \;).
	\end{equation}
When rotating the $\chi^{4}\overline{b_{L}}'s_{R}'$ interaction into the mass basis, one also obtains a $\chi^{4}\overline{s_{L}}d_{R}$ interaction suppressed by a factor $\sim \epsilon_{s}^{3}$. Taking into account the limits on the relevant Wilson coefficient, one finds 
	\begin{equation}
	\Lambda_{\chi} \gtrsim  700 \; \mathrm{GeV} \qquad ( \; \text{from} \; \mathrm{Im}[C_{4}^{sd}] \;).
	\end{equation}
We have also checked that the higher loop order diagrams for the Wilson coefficients do not constrain $\Lambda_{\chi}$ more stringently for $Q_{\rm FN}(X)=-1/2$.

\section{Effects of higher dimensional terms in the scalar potential}
\label{sec:dimension6} 

\begin{figure}[t]
\begin{center}
\includegraphics[width=220pt]{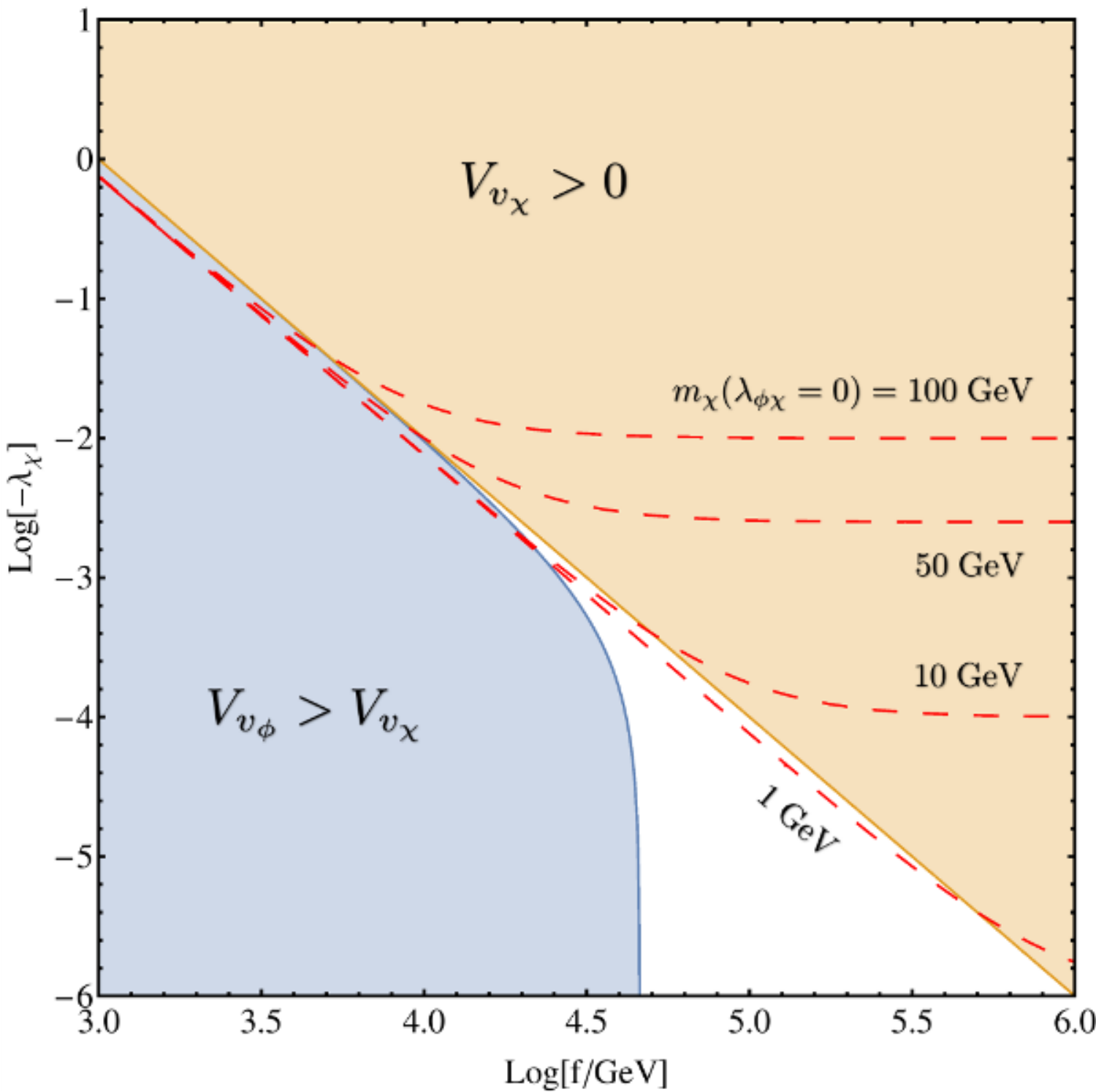}
\end{center}
\caption{\small Parameter space for model B including the higher dimensional operator $\chi^{6}/f^{2}$ and choosing $\lambda_{\chi}$ to be negative. The shaded areas are disallowed as inequality~(\ref{eq:mincond}) is violated. Large contributions to $m_{\chi}$ (shown by the dashed lines)  can only be achieved by a fine tuning between the quartic coupling $\lambda_{\chi}$ and the $\chi^6$ interaction controlled  by $f$. The additional free parameter $\lambda_{\phi \chi}$ tends to give the dominant contribution to $m_{\chi}^{2}$ as in the scenario without $f$.}
\label{fig:parameterspace}
\end{figure}

In this section we look whether any modification to our scalar potential will affect our conclusions.
We illustrate this in the context of Models B where the arguments can be simply explained, however we expect similar statements in Models A.

The main conclusion of this work is that a light scalar, either $\sigma$ or $\chi$ is needed.
To show that this general conclusion does not depend on the specific form of the scalar potential, 
we  add a higher dimensional operator to the potential of Model B-1,
	\begin{align}
	\label{eq:potsimple2}
	V  & = \frac{ \mu_{\phi}^{2} }{ 2 }\phi^{2}+ \frac{ \lambda_{\phi} }{ 4 } \phi^{4} + \frac{ \lambda_{\phi \chi } }{ 4 } \phi^{2}\chi^{2} + \frac{ \mu_{\chi}^{2} }{ 2} \chi^{2}+\frac{ \lambda_{\chi} }{ 4 }\chi^{4} +\frac{1}{8f^{2}}\chi^{6}.
	\end{align}
We want the flavon to have a VEV at the Froggatt--Nielsen scale before EW symmetry breaking, $v_{\chi}=\Lambda_{\chi}$, and zero VEV after EW symmetry breaking. This sets the relations
	\begin{align}
	\mu_{\chi}^{2}+\lambda_{\chi}v_{\chi}^{2}+\frac{3v_{\chi}^{4}}{4f^{2}}=0, \\
	\mu_{\phi}^{2}+\lambda_{\phi}v_{\phi}^{2}=0.
	\end{align}
(from now it is understood that $v_{\chi}=\Lambda_{\chi}$ denotes the initial flavon VEV). An additional condition to obtain the VEV variation is that the EW minimum is deeper than the $v_{\chi}$ minimum and similarly, the $v_{\chi}$ minimum should be deeper than the stationary point at the origin. This gives a condition
	\begin{equation}
	0 > -\lambda_{\chi}v_{\chi}^{4}-\frac{1}{f^{2}}v_{\chi}^{6} > -\lambda_{\phi}v_{\phi}^{4}. \label{eq:mincond}
	\end{equation}
This shows that, outside of fine tuning between the $\lambda_{\chi}$ and $f^{2}$ parameters, both terms wedged in the inequality are at most $\sim\mathcal{O}(\lambda_{\phi}v_{\phi}^{4})$, \emph{i.e.} EW scale. The mass of the flavon is given by
	\begin{align}
	m_{\chi}^{2}  = \mu_{\chi}^{2}+\frac{\lambda_{\phi \chi}v_{\phi}^{2}}{2}  
		      =-\lambda_{ \chi}v_{\chi}^{2}-\frac{3v_{\chi}^{4}}{4f^{2}}+\frac{\lambda_{\phi \chi}v_{\phi}^{2}}{2}. \label{eq:mass}
	\end{align}
The first and third terms can give positive contributions and their sum must be larger than the negative contribution from the second term. Avoiding excessive fine tuning in eq.~(\ref{eq:mincond}) implies that the first two terms in eq.~(\ref{eq:mass}) are at most $\sim\mathcal{O}(\lambda_{\phi}v_{\phi}^{2})$. Hence $m_{\chi}$ is at most EW scale. The situation is illustrated in figure~\ref{fig:parameterspace}, which shows the parameter space in the $\lambda_{\chi}$---$f$ plane. Achieving a contribution to $m_{\chi}\sim10$ GeV requires already quite a bit of tuning between $\lambda_{\chi}$ and $f$.

We now ask if our argument can be generalised to incorporate other possible dimension six operators. The obvious possibilities are
	\begin{equation}
	\phi^{2}\chi^{4}, \quad \phi^{4}\chi^{2}, \quad \phi^{6}.
	\end{equation}
\begin{itemize}
\item The $\phi^{2}\chi^{4}$ contributes neither to the potential at the two minima, nor to the mass term, and therefore has no effect on our above conclusions.
\item The $\phi^{4}\chi^{2}$ term does not contribute to the potential at the two minima, but it contributes a factor $\sim v_{\phi}^{4}/f_{2}^{2}$ to $m_{\chi}^{2}$. Again, this is at most EW scale, if we demand the UV completion scale $f_{2}$ to lie above the EW scale. Hence our conclusions remain unchanged.
\item The $\phi^{6}$ term does not contribute to $m_{\chi}^{2}$ but it does contribute to $V$ at the EW minimum. It can make the EW minimum deeper and therefore alleviate tuning for larger choices of $|\lambda_{\chi}v_{\chi}^{4}|$ and $v_{\chi}^{6}/f^{2}$. However, its contribution to $V$ is $\sim v_{\phi}^{6}/f_{3}^{2}$, which is again at most the EW scale, and this term does not change our conclusions.
\end{itemize}

We expect similar conclusions to hold for model A. However, because the flavon in model A retains a VEV at the EW minimum, it is more difficult to provide a clear analytical argument  showing that higher dimensional operators in the potential do not help increasing the flavon. We have not  found areas of parameter space, when including higher dimensional operators ,which give a large flavon mass, $m_{\sigma} \gtrsim v_{\phi}^{2}/\Lambda_{s}$. Again this can be understood as a consequence of requiring a flat potential in the $\sigma$ direction, in order for the flavon VEV to change by a value of order $\sim \Lambda_{s}$, when the Higgs acquires a VEV of order $\sim v_{\phi}$.

\section{Exact $Z_2$ Symmetry: ruled out by Dark Matter constraints}
\label{sec:darkmatterconstraints}

\begin{figure}[t]
\begin{center}
\includegraphics[width=200pt]{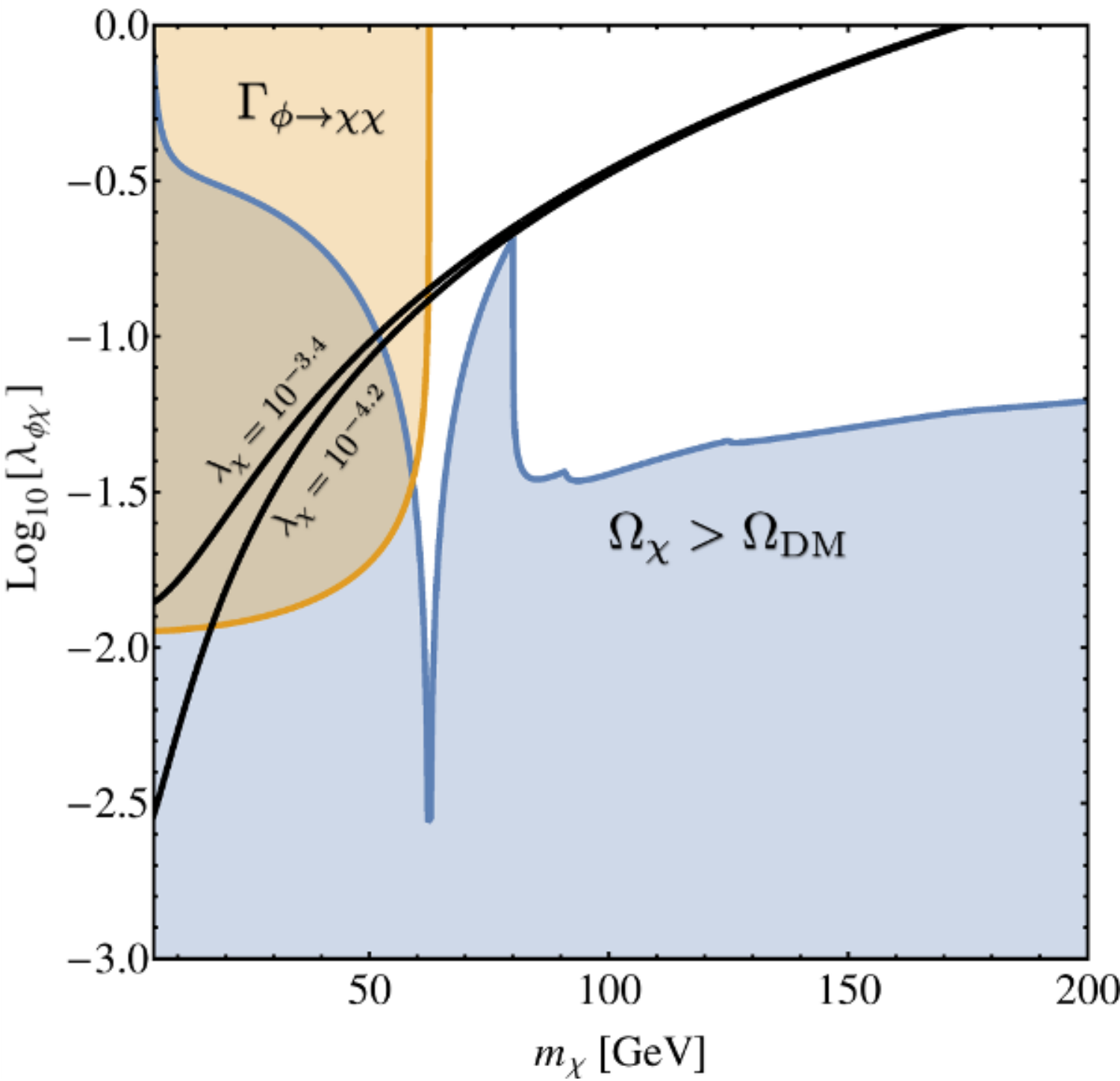}
\includegraphics[width=200pt]{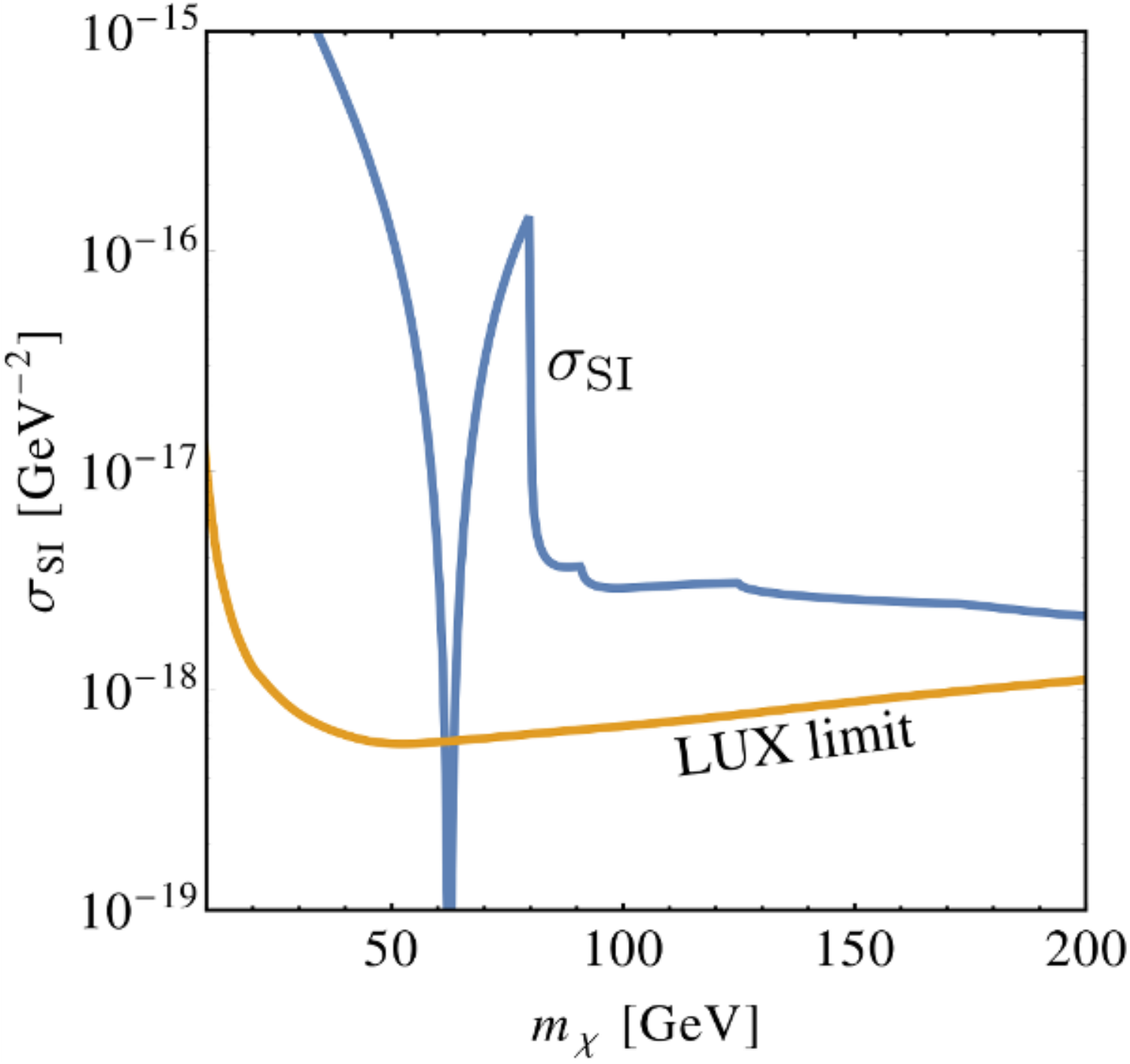}
\end{center}
\caption{\small Left: the black lines show eq.~(\ref{eq:modelbmass}), the relation between $m_{\chi}$ and $\lambda_{\phi \chi}$ for model B. The upper line corresponds to $\lambda_{\chi}=10^{-3.4}$ and the lower line to $\lambda_{\chi}=10^{-4.2}$, which approximately spans the parameter space leading to strong first-order phase transitions. The red region is ruled out by the Higgs signal strength measurement as described in section~\ref{sec:exohiggs}. The blue region is ruled out by the measured DM abundance in the exact $Z_{2}$ case. Right: The approximate spin independent $\chi$-nucleon scattering cross section for model B in the exact $Z_{2}$ case, areas above the yellow line are ruled out by LUX (2016)~\cite{Akerib:2015rjg,Manalaysay}.}
\label{fig:modelB_DM}
\end{figure}
In Model B-1, there is a $Z_2$ symmetry associated with the scalar potential of the  second FN field $\chi$.  
If the $Z_{2}$ is exact then $\chi$ is stable and we show here that this leads to unacceptable dark matter phenomenology. The leading annihilation channel for $\chi$ is through the Higgs portal~\cite{Silveira:1985rk}, which is by now highly constrained~\cite{Djouadi:2011aa,Cline:2013gha,Cheung:2012xb,Duerr:2015aka}. Cross sections may be found in ref.~\cite{Duerr:2015aka}. In our discussion we use the approximation $s \approx 4m_{\chi}^{2}$ at freeze out. The required relation between $\lambda_{\phi \chi}$ and $m_{\chi}$ for Higgs portal DM is met by eq.~(\ref{eq:modelbmass}) at only one point, at $m_{\chi} \approx 50$ GeV, as shown in figure~\ref{fig:modelB_DM}. (The resonance region is slightly broader than in our approximation when the thermal average of the cross section is properly taken into account. The peak just below the $W$ mass is also lower if the thermal average is performed and misses the model B $\lambda_{\phi \chi}$ -- $m_{\chi}$ relation by a greater extent than what is shown in figure~\ref{fig:modelB_DM}~\cite{Duerr:2015aka}). This point is ruled out by constraints from direct detection~\cite{Akerib:2015rjg,Manalaysay} and  the limit on the Higgs width~\cite{ATLAS-CONF-2015-044}. Indeed apart from the resonant regime --- in which our model will lead to a subdominant contribution of $\chi$ to the DM abundance --- the model with an exact $Z_2$ symmetry is ruled out up to a mass $m_{\chi} \approx 380$ GeV. For masses away from the resonance and below 380 GeV the $\chi$ abundance either overcloses the universe or leads to DM-nucleon scattering above the direct detection constraint. (The $\chi$-nucleon scattering cross section scales as $\lambda_{\phi \chi}^{2}$, while the abundance scales as $\lambda_{\phi \chi}^{-2}$, so an approximate estimate of the region excluded by direct detection can be found by finding mass regions where the usual Higgs portal DM is still allowed~\cite{Cline:2013gha}, this is shown on the right panel of figure~\ref{fig:modelB_DM}.) For much of the parameter space we either require a breaking of the $Z_2$ symmetry or new annihilation channels for $\chi$ in order to avoid overclosure. In our analysis we only retain the case with explicit breaking of the $Z_{2}$ symmetry.

\bibliographystyle{JHEP}   
\bibliography{fnewbg} 
\end{document}